\documentclass[11pt]{article}

\usepackage[T1]{fontenc}
\usepackage[utf8]{inputenc}
\usepackage[english]{babel}
\usepackage[a4paper,margin=2.5cm]{geometry}
\usepackage{newtxtext,newtxmath}
\usepackage{amsmath}
\usepackage{graphicx}
\usepackage{float}
\usepackage{array}
\usepackage{xcolor}
\usepackage{braket}
\usepackage[version=3]{mhchem}
\usepackage[super,sort&compress,comma]{natbib}
\usepackage[nolist]{acronym}
\usepackage[colorlinks=true,allcolors=blue]{hyperref}

\newcommand{\supp}{Supporting Information}

\renewcommand{\vec}[1]{\mathbf{#1}}
\newcommand{\anna}[1]{\hat a_{#1}}
\newcommand{\crea}[1]{\hat a_{#1}^\dagger}
\newcommand{\annb}[1]{\hat b_{#1}}
\newcommand{\creb}[1]{\hat b_{#1}^\dagger}

\renewcommand{\P}{\hat P^{(1)}}
\newcounter{statement}
\newcommand{\orba}[1]{\varphi^A_{#1}}
\newcommand{\orbb}[1]{\varphi^B_{#1}}
\newcommand{\wf}[2]{\Psi^{#1}_{#2}}
\newcommand{\sab}[1]{S^{AB}_{#1}}
\newcommand{\sba}[1]{S^{BA}_{#1}}

\newcommand{\elst}{\ensuremath{V^{\text{elst}}\,}}
\newcommand{\Sexch}{\ensuremath{X^\text{S}}\,}
\newcommand{\exch}{\ensuremath{V^{\text{exch}}\,}}

\makeatletter
\renewcommand{\@maketitle}{%
	\newpage
	\null
	\vskip 2em%
	\begin{center}%
		{\LARGE \@title \par}%
		\vskip 1.5em%
		{\large
			\lineskip .5em%
			\@author
			\par}%
		\vskip 1em%
	\end{center}%
	\par
	\vskip 1.5em%
}
\makeatother

\title{On the importance of multi-configurational and exchange effects in molecular aggregates}
\author{Andy Kaiser$^{a,b}$, Francesco Aquilante$^c$, Patrick Staschick$^a$, Oliver K\"uhn$^a$, Sergey I. Bokarev$^{d,a}$\\[0.8em]	
\small $^a$Institut f\"ur Physik, Universit\"at Rostock, Albert-Einstein-Str. 23--24, 18059 Rostock, Germany\\
\small $^b$Leibniz-Institut f\"ur Katalyse e.V. (LIKAT), Albert-Einstein-Str. 29a, 18059 Rostock, Germany\\
\small $^c$Theory and Simulation of Materials (THEOS), and National Centre for Computational Design and Discovery of Novel Materials (MARVEL),\\
\small \`Ecole Polytechnique F\'ed\'erale de Lausanne, CH-1015 Lausanne, Switzerland\\
\small $^d$Chemistry Department, School of Natural Sciences, Technical University of Munich, Lichtenbergstr. 4, 85748 Garching, Germany\\
\small Corresponding author: \texttt{sergey.bokarev@tum.de}}
\date{}

\begin{document}
\maketitle

\begin{abstract}
We present an extension of the Frenkel exciton model to incorporate exchange interactions between monomers in molecular aggregates in conjunction with a multi-reference electronic structure approach. Our derivation, which combines the Frenkel exciton Hamiltonian and the single-electron pair exchange approximation, yields a non-perturbative, variational expression for the exchange coupling that naturally excludes any basis set superposition error. The method has been implemented in OpenMolcas and enables combination with multi-reference electronic structure techniques. The main objective of the present study is to assess the role of exchange in systems with strong multi-configurational character. Illustrative examples demonstrate how the inclusion of exchange at different levels of approximation can substantially alter the magnitude and sign of intermonomer couplings and thus, for instance, potentially converting the predicted classification of the aggregate from H-type to J-type. Comparison with TDDFT-based couplings highlights significant discrepancies arising from multi-reference effects, double excitations, and Rydberg transitions. Overall, this approach advances the predictive modeling of photophysical and photochemical processes in aggregates of polyacenes, carotenoids, and other systems where multi-configurational and Rydberg states are essential.
\end{abstract}

\begin{acronym}
    \acro{AO}{Atomic Orbital}
    \acro{BChl A}{Bacteriochlorophyll A}
    \acro{BSSE}{Basis Set Superposition Error}
    \acro{CASPT2}{Complete Active Space Second-order Perturbation Theory}
    \acro{CASSCF}{Complete Active Space Self-Consistent Field}
    \acro{CHA}{Chemical Hamiltonian Approach}
    \acro{CI}{Configuration Interaction}
    \acro{DFT}{Density Functional Theory}
    \acro{DMRG}{Density Matrix Renormalization Group}
    \acro{FOD}{Formaldehydeoxime}
    \acro{LH2}{Light-harvesting complex 2}
    \acro{MO}{Molecular Orbital}
    \acro{RASSCF}{Restricted Active Space Self-Consistent Field}
    \acro{RASPT2}{Restricted Active Space Second-order order Perturbation Theory}
    \acro{RG1}{Rhodopin $\beta$-D-glucoside}
    \acro{SAC}{Symmetry-Adapted Cluster}
    \acro{SAPT}{Symmetry-Adapted Perturbation Theory}
    \acro{TDDFT}{Time-dependent Density Functional Theory}
\end{acronym}


\section{Introduction}
\label{sec:intro}

Understanding molecular aggregates and exciton dynamics is crucial to unraveling
photophysics and photochemistry, transport properties, and rational design of materials and molecular devices, such as organic light emitters and photovoltaics.\cite{Ye_AM_2015,Gao_A_2021,Mas-Montoya_AFM_2017,Feng_JPCC_2022,Xu_APL_2024}
Further, exciton dynamics are at the heart of photosynthetic light harvesting, realized by coupled chromophores in pigment-protein complexes.\cite{blankenship14_,renger01_137}
Theoretical calculations offer indispensable help in interpreting experimental results and gaining atomistic insight into the photoinduced processes in aggregates.
During decades of theoretical studies, several methods have been introduced, varying strongly in their complexity, computational cost, and accuracy.
For cases where a strong correlation between multiple monomers plays an important role, multimonomeric many-body methods were formulated, including excitonic coupled cluster~\cite{Liu_MP_2019,Dutoi_MP_2019}, embedding \ac{CI} method~\cite{Pitesa_JCTC_2024}, active space decomposition \ac{DMRG},~\cite{Parker_JCP_2013,Parker_JCP_2014a} and other approximate methods.~\cite{Sisto_ACR_2014, Morrison_JPCL_2015} 
Although the many-body expansion of the intermonomer coupling increases the accuracy, modeling of large aggregates remains challenging, which calls for more efficient approaches, limiting it
to the two-body interaction alone.

These simplifications lead to the  Frenkel model (sometimes called Frenkel-Davydov).~\cite{Davidov_UFN_1964,may23,Morrison_JCTC_2014,Morrison_JCP_2024,Nematiaram_CM_2021,Bae_JPCA_2020,Schroter_PR_2015}
Its popularity is related to intermonomer interactions being considered at the pairwise level, where coupled local excitations in monomers form the delocalized excited states similar to the ``true'' aggregate states and serve as a convenient basis for their representation.

The problem {of defining a Frenkel exciton Hamiltonian} can be attacked from two directions: One can perform an ``adiabatic'' supermolecular calculation of a dimer followed by a diabatization scheme. 
However, in this ``top-down'' approach, there is an ambiguity in how to diabatize the problem, and several schemes are in use; see Refs. \citenum{Carreras_JCTC_2019,Tamura_JPCA_2016,Arago_JCP_2015,Liu_JCP_2015} to name but a few.
Moreover, supermolecular calculations might get prohibitively expensive if a high-level electronic structure method is applied.
Finally, such calculations are prone to \ac{BSSE} - a non-physical effect stemming from the incompleteness of the orbital basis - and this error cannot be removed.
From this point of view, following a ``bottom-up'', diabatic-by-construction strategy is advantageous, where monomer excitations and intermonomer couplings are calculated separately.
In both approaches, any electronic structure method can be used to parameterize the Frenkel model, which is limited only by affordability, where the second approach scales more favorably.

In the ``bottom-up'' approach, the intermonomer coupling, in turn, can be calculated at different levels of complexity starting from the transition dipole coupling,~\cite{Egorov_PP_2009,Preusse_SD_2016} coupling of transition densities,~\cite{Krueger_JPCB_1998, Fuckel_JCP_2008, Caricato_JCTC_2015, Hofener_JCTC_2016, Barcza_JCTC_2023, Kaiser_JCTC_2023} 
and even including the exchange effects~\cite{Morrison_JCTC_2014,Li_JCTC_2017} that can be essential in certain cases.~\cite{Bai_FD_2019,Bai_JPCC_2020,Cravcenco_JPCA_2020}
A notable advancement in the field following the latter route is the computational protocol based on linear-response \ac{TDDFT}.~\cite{Hsu_JCP_2001, Iozzi_JCP_2004, Russo_JPCB_2007} 
However, \ac{TDDFT} cannot be used in cases where molecules possess multi-configurational wave functions and/or excited electronic states are of double-excitation character. 
Moreover, the results of \ac{TDDFT} strongly depend on the exchange-correlation functional, with many conventional functionals giving notable errors for (intramonomer) charge-transfer 
and Rydberg states, as well as for excited states of different multiplicities in general\cite{Hirata1999,Hirata_CPL_1999b,Tozer_PCCP_2000,Casida_JCP_1998,Eriksen_MP_2013,Fuks_PRA_2014} and for aggregate systems in particular.~\cite{Bruckner_CP_2017} 
That is why there is a need for a theory with systematic convergence to the exact limit. 

\begin{figure}
	\centering
	\includegraphics[width=1.0\linewidth]{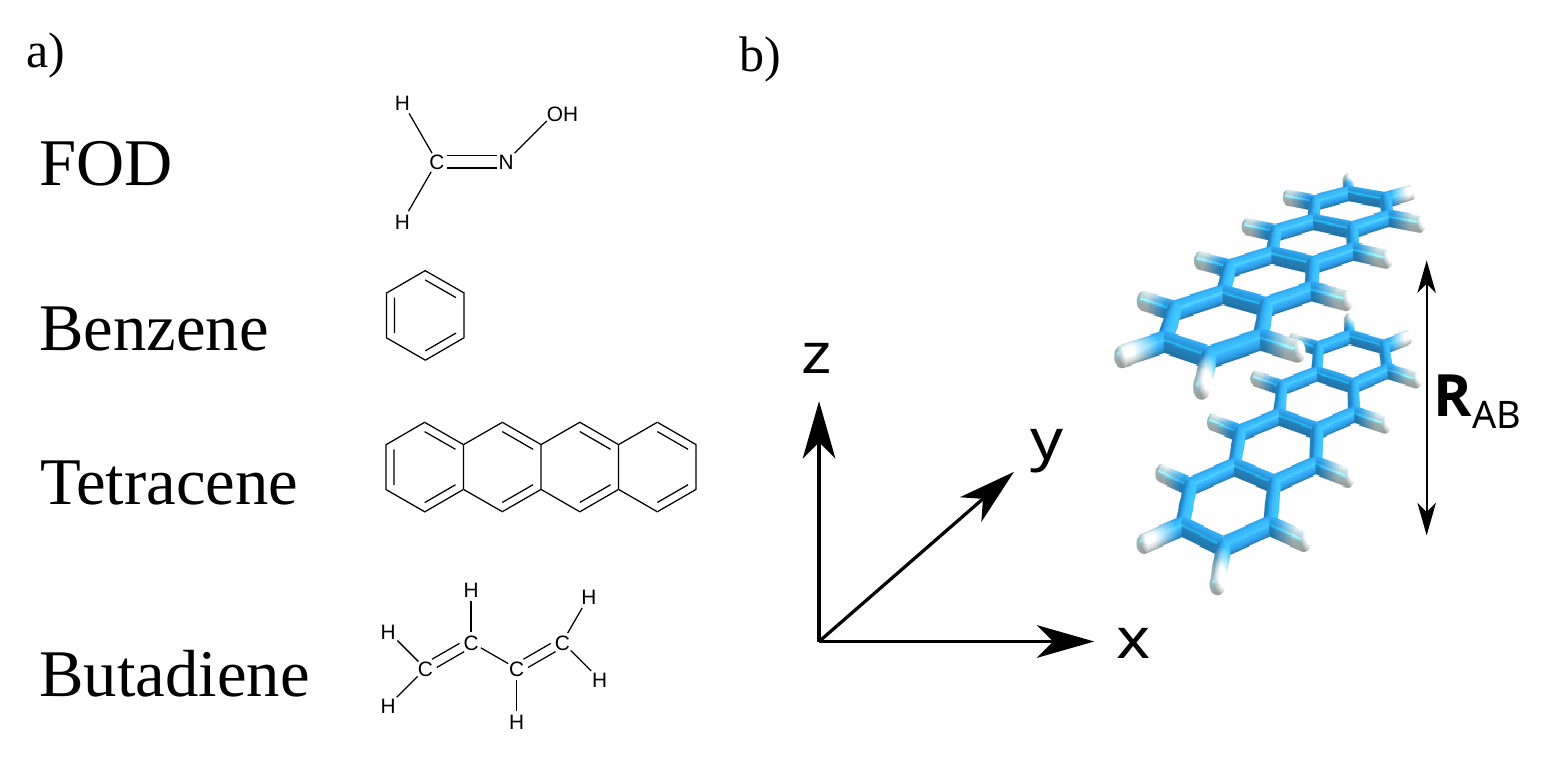}
	\caption{ 
    Model systems studied in this work: a) structural formula of \acf{FOD}, benzene, tetracene and butadiene; b) relative orientation of planar monomers in a dimer. \label{fig:systems}}
\end{figure}

To overcome the limitations of \ac{TDDFT} and other single-reference methods, we extend the protocol introduced in Ref.~\citenum{Kaiser_JCTC_2023}, which is based on the \ac{RASSCF}/\ac{RASPT2}\cite{malmqvistRestrictedActiveSpace1990} electronic structure method, to include couplings that arise from the exchange interaction. Further, we discuss the  implementation of  the protocol into the \texttt{OpenMolcas} program package.~\cite{LiManni_JCTC_2023}
The main focus of this paper is on the importance of multi-configurational effects, the  systematic analysis of the different levels of exchange coupling treatment and the influence of the non-physical \ac{BSSE} contributions.
To this end, the  protocol is applied to  dimers of the following  molecular systems (cf. Fig. \ref{fig:systems}):

\paragraph*{\ac{FOD}}
It has already been studied by us, although  considering only the Coulomb coupling.~\cite{plotz14_174101,Kaiser_JCTC_2023} Due to the presence of C, N and O atoms, a double bond, and a lone pair, it represents one of the smallest model systems having properties similar to heterocycles.
For example, this molecule exhibits low-lying electronic states of Rydberg character that are also typical for heterocycles.\cite{Zilberg_JPCA_2012,Schalk_JCP_2018a}
These Rydberg states may be unreliably predicted by linear-response \ac{TDDFT}.

\paragraph*{Butadiene}
 Butadiene is a prominent example of a monomer possessing a dark second excited state due to its double-excitation character,\cite{Shu_JACS_2017,Dong_JCTC_2019,DoCasal_CS_2023a} thus making the linear-response \ac{TDDFT} protocol inapplicable in this case.  In addition to the multi-configurational double-excitation character, such systems are of interest for the present study as the dipole approximation is inapplicable due to the dark nature of states, and exchange coupling may come to the forefront.

\paragraph*{Benzene and tetracene}
Polyacenes represent attractive materials for organic photovoltaics and are often discussed in the context of singlet fission, which increases the efficiency of solar cells.~\cite{Daiber_JPCL_2020,Daiber_AEL_2021,Wang_JMCA_2023,MacQueen_MH_2018}
There is a plethora of computational studies using \ac{TDDFT}, single-reference and multi-reference wave function methods.~\cite{Hajgato_JCP_2009, Horn_TCA_2014, Filatov_JCP_2014, Kurashige_BCSJ_2014, Bettinger_JCTC_2016, Yang_PNAS_2016, Bettanin_JCTC_2017, Casanova_JCP_2018, C.A.Valente_JCP_2021, Dai_C_2023, Sandoval-Salinas_JCP_2023}
The consensus from these works is that although, with growing size, the electronic structure stays within the applicability limits of single-reference methods, the decreasing weight of the closed-shell configuration in the ground state becomes problematic.
For excited states, the multi-configurational character increases, and the doubly-excited configurations gain importance as the size  increases. 
For instance, these configurations are needed to correctly predict the order of excited states.
Here, we consider benzene as the simplest parent compound for polyacenes and tetracene, which already has a substantially multi-configurational character.

The article is organized as follows. In Sections~\ref{sec:formalism} and \ref{sec:ME}, we present the formalism employed here. Section~\ref{sec:BSSE} discusses the influence of the \ac{BSSE} and identifies terms that need to be excluded in the \ac{BSSE}-free theory. In Section~\ref{sec:comparison}, we establish relations between our approach and other techniques and present the hierarchy to include exchange effects.
The particular implementation into the \texttt{OpenMolcas} program package is briefly described in Section~\ref{sec:protocol}, and the computational details for our test calculations are given in Section~\ref{sec:comp_details}.
The results of the calculations are discussed in Section~\ref{sec:results}, where we assess the validity of different approximations to exchange and discuss the importance of including multi-configurational and exchange effects to study aggregates of organic systems.
The article is summarized in Section~\ref{sec:conclusions}.

\section{Theory}
\label{sec:theory}
\subsection{General framework}
\label{sec:formalism}

We will focus on the purely electronic problem and thus assume a clamped-nuclei approximation.
Consider a dimer as a model for a molecular aggregate with pairwise interaction between the monomers denoted $A$ and $B$. Within the ''bottom-up'' approach, we separate the electronic Hamiltonian into additive monomer terms and intermonomer interactions: 

\begin{align}\label{eq:Ham}
    \hat H_\text{agg} &= \underbrace{\hat h^A\otimes\hat 1^B + \hat 1^A\otimes \hat h^B + \hat V^A_\text{ee}\otimes\hat 1^B + \hat 1^A\otimes\hat V^B_\text{ee} }_{\hat{H}_{\text{intra}}} \nonumber\\
    &+ \underbrace{\hat V^{AB}_\text{en}\otimes\hat 1^B + \hat 1^A\otimes\hat V^{BA}_\text{en}  + \hat V^{AB}_\text{ee}  +  \hat V^{AB}_\text{nn}}_{\hat{H}_{\text{inter}}} \, .
\end{align}
Here, $\hat h^M=\hat T + \hat V^M_\text{en}$ is a one-electron monomer-specific Hamiltonian ($M=A,B$), including the kinetic energy and electron-nuclear interaction, and $\hat V^M_\text{ee}$ is the respective intramonomer electron-electron interaction term. 
$\hat V^{AB}_\text{en}$ and $\hat V^{BA}_\text{en}$ denote the interaction of electrons on monomer $A$ with the nuclei of $B$ and vice versa; $\hat V^{AB}_\text{ee}$ is the intermonomer electron repulsion.
We group electronic terms into the intramonomer part $\hat H_\text{intra}$ and the intermonomer interaction $\hat H_\text{inter}$.
This distinction will be important later for the discussion of the \ac{BSSE}.
$V^{AB}_\text{nn}$ is the nuclear repulsion energy; we will not separate it further into different terms with respect to intra- and intermonomer interactions, as it is not subject to \ac{BSSE}. Note that we do not take into account effects due to an environment. This could be done, for instance, using charge or density embedding.~\cite{Pitesa_JCTC_2024, Barcza_JCTC_2023}

As an ansatz for the dimer's  wave function basis, we consider anti-symmetrized products $\mathcal{A}\{\Psi^A_a\Psi^B_b\}$ of in general non-orthogonal many-electron wave functions $\Psi^A_a$ and $\Psi^B_b$ of isolated monomers. 
In principle, such a basis is complete in the fermionic subspace of the total Hilbert space; in practice, however, only a reduced  monomeric basis can be used. 
In what follows, we consider the manifolds of states having $N_A$ and $N_B$ electrons, respectively; charge-transfer effects between the monomers are excluded.
Since the monomeric wave functions $\Psi^A_a$ and $\Psi^B_b$ remain fixed in our calculations, the formalism should be suitable for cases when the long-range static (strong) intermonomer correlation effects are negligible.
The extent of the inclusion of dynamic (weak) correlation then depends on the chosen dimer basis. 
In any case, the electrons within monomers are, in principle, fully correlated as we treat the respective states locally with a multi-configurational method, e.g., \ac{RASSCF}/\ac{RASPT2}, which can be systematically improved to the full \ac{CI} limit. The accurate account for intramonomer correlations while maintaining the local character of the basis is an advantage of the non-orthogonal CI strategy used in this work. Notice that a non-orthogonal CI approach to calculate couplings  between multi-configurational states of monomers is also followed by the \texttt{GronOR} software.~\cite{Straatsma_JCTC_2022}  
Finally, we point out that we omit the consideration of the spin part of the aggregate wave function. 
Effectively, this means that the maximum combination of spin momenta $S_A+S_B$ with the largest spin projection $M_A+M_B$ on the quantization axis is assumed. 

To account for the exchange interaction, we decompose the intermolecular antisymmetrizer~\cite{Landman_CPL_1971} in terms of permutations of one, two, \ldots, $n$ electron pairs between the monomeric wave functions, corresponding to operators $\P$ $\hat P^{(2)}$, \ldots, $\hat P^{(n)}$: 
\begin{align}\label{eq:antisym_WF}
	\mathcal{A}\{\Psi^A_a\Psi^B_b\} = &\sqrt{\frac{N_A!N_B!}{(N_A+N_B)!}}(\hat 1+\P + \ldots + \hat P^{(n)}) \nonumber\\ 
    &\times\Psi^A_a(1,\ldots,N_A) \Psi^B_b(N_A+1,\ldots,N_A+N_B)\, ,
\end{align}
where $n=\min\{N_A, N_B\}$.
The unity operator here and in the following should be understood as $\hat 1=\hat 1^A\otimes\hat 1^B$.
The intermolecular permutation operators  can be shown to be proportional to the powers of orbital overlaps between monomers,~\cite{Tyrcha_JCP_2024,Stone2013} i.e., $\hat P^{(i)} \propto S^{2i}$. 
In what follows, the central assumption will be that the wave function overlap is small enough to justify approximating the antisymmetrizer by the term $\hat1+\P$, i.e., up to order $S^2$. This approximation will be further discussed in  Section~\ref{sec:validity_P1}.

Since this approximation is heavily used in \ac{SAPT},~\cite{Patkowski_WCMS_2020} it is important to note the difference: In the present approach, we  proceed in the spirit of the \ac{CI} method and write a Hamiltonian matrix in the basis of $\mathcal{A}\{\Psi^A_a\Psi^B_b\}$ states with a subsequent diagonalization. Thus, we are  staying variational, which is  in contrast to perturbation theory. 
Since we are not constructing a perturbation series, there is no need to introduce a zero-order 
Hamiltonian and we can directly work with non-orthogonal basis states and a non-Hermitian Hamiltonian, performing a posteriori orthogonalization; see Eq.~\eqref{eq:gen_eigv}.
In that sense, our approach is similar to aggregate \ac{CI} of Pite\v sa et al.~\cite{Pitesa_JCTC_2024} but with a different treatment of exchange effects.
Our approach still partially resembles \ac{SAPT}, although commonly only expectation values in the ground state are in focus in the latter  (for a more detailed discussion see Sec. \ref{sec:comparison}).
Moreover, the electronic structure methods of choice for monomers are usually of single-reference character, with the exception of Ref.~\citenum{Hapka_JCTC_2021}, which employs multi-reference methods and allows for general computation of energy expectation values in excited states. 
In contrast, here, we are interested in transitions between electronic states of monomers, building a basis for a Frenkel exciton model, being extendable beyond the dimer. 

To derive the form of Hamiltonian matrix elements, we employ the second quantization formalism.
First, we assume that the orbital bases of $A$ and $B$ are complete and thus span the same (full) space.
It allows us to exchange the ladder operators of $A$ (denoted as $\crea{}$ and $\anna{}$) and $B$ ($\creb{}$ and $\annb{}$) in the second quantization formalism, for example,
\begin{equation}\label{eq:a2b}
	\creb{j}=\sum_k S^{AB}_{kj}
    \crea{k} \quad \text{and} \quad \crea{i}=\sum_l S^{BA}_{li}
    \creb{l} \, ,
\end{equation}
where $\sab{kj}=\braket{\varphi^A_{k}|\varphi^B_{j}}$ and $\sba{li}=\braket{\varphi^B_{l}|\varphi^A_{i}}$ are overlaps between the  monomeric \acp{MO}  $\varphi^M (M=A,B)$.
The $\P$ operator can therefore be represented in second quantization as \cite{Tyrcha_JCP_2024}
\begin{equation}\label{eq:P1}
	\hat P^{(1)} = -\sum_{\substack{kn\in A\\lm\in B}} 
    \sab{km}\sba{ln}\
    \crea{k}\anna{n}\otimes\creb{l}\annb{m} \, .
\end{equation}
The Hamiltonian matrix elements, Eq.~\eqref{eq:Ham}, are obtained
assuming that the electrons of $A$ and $B$ are distinguishable, i.e., $\crea{}$ and $\anna{}$ operators act only on wavefunctions of $A$ and respectively for $B$  as is already reflected in the form of the $\P$ operator, Eq.~\eqref{eq:P1}.
This corresponds to the usual anticommutation algebra for operators within monomers, whereas operators of different monomers commute~\cite{Rybak_JCP_1991}
\begin{equation}\label{eq:commutation}
    [\anna{i},\annb{j}]=[\anna{i},\creb{j}]=[\crea{i},\annb{j}]=[\crea{i},\creb{j}]=0 \, .
\end{equation}
%

\subsection{Matrix elements}
\label{sec:ME}

The one- and two-electron Hamiltonian terms, e.g., for monomer $A$, can be written in second quantization as
\begin{align}
    \hat O_1^A &= \sum_{pq\in A} \bra{\orba{p}}\hat o_1^A\ket{\orba{q}}\crea{p} \anna{q} \otimes \hat 1^B\\
    \hat O_2^A &= \frac{1}{2}\sum_{pqrs \in A} \bra{\orba{p}\orba{q}}\hat o_2^A\ket{\orba{r}\orba{s}}\crea{p} \crea{q} \anna{s} \anna{r} \otimes \hat 1^B \, ;
\end{align}
for monomer $B$ and non-separable terms, the orbitals and operators must be changed accordingly.
Further, 
\begin{equation*}
    \hat V_\text{ee}^{AB}=\sum_{\substack{pr\in A\\qs \in B}} \bra{\orba{p}\orbb{q}}\hat o_2^{AB}\ket{\orba{r}\orbb{s}}\crea{p}   \anna{r} \otimes \creb{q}\annb{s} \, ;
\end{equation*}
note the absence of the factor 1/2 in this case.
Here and in the following, the indices $p,q,r,s,\ldots$ denote spin orbitals.

The overlap between basis states corresponds to the matrix elements of the operator $(\hat1+\P)^2$ between simple Hartree products of monomer states (note that the truncated operator $\hat1+\P$ is not in general idempotent in contrast to the full antisymmetrizer $\mathcal A$)
%
\begin{align}\label{eq:ovlp_neutral}
    \braket{\wf{A}{a}\wf{B}{b}|\mathcal{A}^{\dagger}\mathcal{A}|\wf{A}{c}\wf{B}{d}} &\approx \bra{\wf{A}{a}\wf{B}{b}}(\hat 1 + \P)^2\ket{\wf{A}{c}\wf{B}{d}} \nonumber\\
    &\approx \bra{\wf{A}{a}\wf{B}{b}}\hat 1 + 2\P\ket{\wf{A}{c}\wf{B}{d}} \nonumber\\
    &= \delta_{bd} \delta_{ac} - 2\sum_{\substack{kn\in A\\lm\in B}}\sab{km}\sba{ln}
    \gamma^{ac}_{kn}\gamma^{bd}_{lm} \, .
\end{align}
Here, we have introduced one-electron transition density matrices with elements $\gamma^{ac}_{kn}=\bra{\wf{A}{a}}\crea{k}\anna{n}\ket{\wf{A}{c}}$ and $\gamma^{bd}_{lm}=\bra{\wf{B}{b}}\creb{l}\annb{m}\ket{\wf{B}{d}}$ and applied the $S^2$ approximation.

Next, we consider the nuclear repulsion term, $\hat V^{AB}_\text{nn}$. 
Due to the  non-orthogonality of monomer basis states, its matrix representation has two terms, i.e., one due to electrostatics and one due to overlap
\begin{align}\label{eq:Vnn}
    \bra{\wf{A}{a}\wf{B}{b}}\mathcal{A}^{\dagger}\hat V^{AB}_\text{nn}&\mathcal{A}\ket{\wf{A}{c}\wf{B}{d}}   \nonumber\\
    &\approx V^{AB}_\text{nn}\delta_{bd} \delta_{ac} - 2V^{AB}_\text{nn}\sum_{klmn}\sab{km}\sba{ln}
    \gamma^{ac}_{kn}\gamma^{bd}_{lm} \, .
\end{align}

For the one-electron terms $\hat h^A\otimes\hat 1^B + \hat V^{AB}_\text{en}\otimes\hat 1^B$ of monomer $A$ in Eq.~\eqref{eq:Ham}, one has
\begin{subequations}
\begin{align} 
    \bra{\wf{A}{a}\wf{B}{b}}&[(\hat h^A+\hat V^{AB}_\text{en})\otimes\hat 1^B](\hat 1+\P)\ket{\wf{A}{c}\wf{B}{d}} \nonumber\\
    &=\delta_{bd}\sum_{pq\in A} \big[h_{pq}^A +  (V^{AB}_\text{en})_{pq}\big]\gamma^{ac}_{pq}  \label{eq:1eA_noS}\\
    &{-\sum_{\substack{pn\in A\\lm\in B}}  \sba{ln}\gamma^{bd}_{lm}}\Bigg(
    \big[\underbrace{h_{pm}^A}_\text{BSSE}  + (V^{AB}_\text{en})_{pm}\big] \gamma^{ac}_{pn} \label{eq:1eA_S}\\
    &- \sum_{qk\in A}\sab{km}\big[\underbrace{h_{pq}^A}_\text{BSSE} +  (V^{AB}_\text{en})_{pq}\big]\gamma^{ac}_{pkqn}\Bigg) \label{eq:1eA_S2}\, .
\end{align}
\end{subequations}
Here, $\gamma^{ac}_{pkqn}=\bra{\wf{A}{a}}\crea{p}\crea{k}\anna{q}\anna{n}\ket{\wf{A}{c}}$ is an element of the two-particle transition density matrix. 
Further, $h_{pq\in A}=\bra{\orba{p}}\hat h \ket{\orba{q}}$ and $h_{p\in A, m\in B}=\bra{\orba{p}}\hat h \ket{\orbb{m}}$; and analogously for matrix elements of $\hat V^{AB}_\text{en}$.
The one-electron terms for $B$ can be obtained analogously, changing all the quantities of monomer $A$ to $B$ and vice versa. 
Here, we assumed the completeness of the orbital basis $\sum_k \ket{\varphi^M_k}\bra{\varphi^M_k}=\hat 1^M, (M=A,B)$.

The two-electron intramonomer terms for $A$ read 
\begin{subequations}
    \begin{align}
    &\bra{\wf{A}{a}\wf{B}{b}}[\hat V^A_\text{ee}\otimes\hat 1^B](\hat 1+\P)\ket{\wf{A}{c}\wf{B}{d}} \nonumber\\
    &= \frac{1}{2} \delta_{bd}\sum_{pqrs\in A} g_{pqrs} 
    \gamma^{ac}_{pqsr} \label{eq:2eA_noS}\\
    &-\frac{1}{2}\sum_{\substack{pqrn\in A\\lm\in B}} \sba{ln}\gamma^{bd}_{lm} {\bigg[}\gamma^{ac}_{pqrn}\Big(g_{pqmr}-g_{pqrm}\Big) \label{eq:2eA_S}\\ 
    &+ \sum_{ks\in A} \sab{km} g_{pqrs}\gamma^{ac}_{pqksrn}\bigg] \label{eq:2eA_S2}
\end{align}
\end{subequations}
with the two-electron integrals defined as $g_{pqrs\in A}=\braket{\orba{p}\orba{q}|\orba{r}\orba{s}}$ in the Dirac notation.  The matrix elements for $\hat 1^A\otimes\hat V^B_\text{ee}$ can be derived analogously.

Finally, the two-electron intermonomer interaction term gives 
\begin{subequations}
\begin{align}
    \bra{\wf{A}{a}\wf{B}{b}}&\hat V^{AB}_\text{ee}(\hat 1+\P)\ket{\wf{A}{c}\wf{B}{d}}  \nonumber\\
    &=\sum_{\substack{pr\in A\\ qs \in B}}\gamma^{ac}_{pr}\gamma^{bd}_{qs} \Big[g_{pqrs}-g_{pqsr}\Big] \label{eq:2eAB_noS}\\
    &+ \sum_{\substack{pn\in A\\ qm \in B}}\bigg[\sum_{ls\in B} \sba{ln} 
    g_{pqms}
    \gamma^{ac}_{pn}\gamma^{bd}_{qlsm} \label{eq:2eBA_S}\\
    &+ \sum_{kr\in A} \sab{km} 
    g_{pqrn}
    \gamma^{ac}_{pkrn}\gamma^{bd}_{qm} \label{eq:2eAB_S}\\
    &-  \sum_{ls\in B} \sab{km}\sba{ln} 
    g_{pqrs}
     \gamma^{ac}_{pkrn}\gamma^{bd}_{qlsm}\bigg] \label{eq:2eAB_S2}
\end{align}
\end{subequations} 
Analogous expressions for the matrix elements of the $\P\hat O$ operators can be found in the \supp, Sec.~S1.

Further, one can combine the $\hat h$ part of Eq.~\eqref{eq:1eA_noS} and~\eqref{eq:2eA_noS} to obtain the energies of the monomeric basis states
\begin{align}\label{eq:energy}
    E^M_i=\delta_{ij}\sum_{pq\in M}\Big[ h_{pq}\gamma^{ij}_{pq} + \frac{1}{2}\sum_{rs \in M} g_{pqrs}
    \gamma^{ij}_{pqsr}\Big] \,,
\end{align}
forming the diagonal of the Hamiltonian matrix.
The interaction energies of the static electron densities ($a=c$ and $b=d$) with the nuclei of the other monomer also populate the diagonal. 
All other terms constitute the off-diagonal elements; note that the interaction of the transition density with the nuclear charge of the other monomer, e.g., $\delta_{bd}\sum_{pq}(\hat V^{AB}_\text{en})_{pq}\gamma^{ac}_{pq}$ for $A$, also contributes to the off-diagonal coupling. 

Using properly normalized partially antisymmetrized basis states, a general matrix element of the aggregate Hamiltonian has the following form 
\begin{align}
     H_{abcd}=\frac{\bra{\wf{A}{a}\wf{B}{b}}(\hat1+\P)\hat H_\text{agg}(\hat1+\P)\ket{\wf{A}{c}\wf{B}{d}}}{\sqrt{\bra{\wf{A}{a}\wf{B}{b}}(\hat1+\P)^2\ket{\wf{A}{a}\wf{B}{b}}\bra{\wf{A}{c}\wf{B}{d}}(\hat1+\P)^2\ket{\wf{A}{c}\wf{B}{d}}}}\, .
\end{align}
This expression can be expanded  in accordance with our $S^2$ approximation, resulting in
\begin{align}\label{eq:normH}
     H_{abcd}
     &\approx\bra{\wf{A}{a}\wf{B}{b}}\hat H_\text{agg}\ket{\wf{A}{c}\wf{B}{d}}+\bra{\wf{A}{a}\wf{B}{b}}\P \hat H_\text{agg}\ket{\wf{A}{c}\wf{B}{d}} \nonumber \\
     &+\bra{\wf{A}{a}\wf{B}{b}}\hat H_\text{agg}\P\ket{\wf{A}{c}\wf{B}{d}} +N_{abcd} \,.
\end{align}
The last term stems  from the normalization and reads 
\begin{align}\label{eq:normcorr}
   N_{abcd} &=- \big[\bra{\wf{A}{a}\wf{B}{b}}\P \ket{\wf{A}{a}\wf{B}{b}} \nonumber
   \\ 
   &+\bra{\wf{A}{c}\wf{B}{d}}\P\ket{\wf{A}{c}\wf{B}{d}}\big]\bra{\wf{A}{a}\wf{B}{b}}\hat H_\text{agg}\ket{\wf{A}{c}\wf{B}{d}}\,.
\end{align}
It will be referred to as {normalization correction} (or $N$) in the following.

\subsection{Eliminating  \ac{BSSE} terms}
\label{sec:BSSE}

Notice that the starting point for the derivation of expressions for the matrix elements in Section~\ref{sec:formalism} was the completeness of the monomeric orbital basis, providing freedom to choose orbitals of either of the monomers, Eqs.~\eqref{eq:a2b}.     
In practice, however, one uses finite (incomplete) \ac{AO} bases on monomers; see Sec.~\ref{sec:validity_P1}.
Hence, special care of \ac{BSSE} must be taken.~\cite{Gianinetti_IJQC_1996, Stone2013,Kaiser_JCTC_2023} 

Another concern about \ac{BSSE} is connected to the fragmentation approach, i.e., the fact that we start from the dimer Hamiltonian naturally containing the terms whose matrix elements are classified as \ac{BSSE}, once the Hamiltonian is separated  into monomeric and coupling contributions.

To avoid  \ac{BSSE}, we use the idea of the \ac{CHA}~\cite{Mayer_IJQC_1998} and briefly review the argument behind it.
When considering an {intra}molecular operator acting {on a respective monomer orbital, }
its action, e.g., for the one-electron Hamiltonian of $A$ can be written as follows:
\begin{equation}\label{eq:CHA}
    \hat h^A \orba{i} = \hat P^A \hat h^A \orba{i} + (\hat 1-\hat P^A)\hat h^A \orba{i} \, ,
\end{equation}
where the operator $\hat P^A = \sum_{i\in A}\ket{\orba{i}}\bra{\orba{i}}$ projects onto the subspace spanned by the orbitals of $A$.
For finite basis sets on monomers, the contribution of the complementary $\hat 1-\hat P^A$ part can be substantial.
Although mathematically it must be included, its physical sense is the contribution to the \ac{BSSE}, which should be eliminated to produce reliable interaction energy estimates for states and couplings for the electronic transitions. 
Similar reasoning can also be applied to two-electron integrals.~\cite{Mayer_IJQC_1998}
Effectively, this means that mixed terms including overlap, one- and two-electron integrals with \acp{MO} from different monomers need to be neglected when dealing with intramolecular operators- denoted $\hat H_\text{intra}$ in Eq.~\eqref{eq:Ham}  - while the intermolecular terms $\hat H_\text{inter}$ stay intact.
It means that, in the \ac{BSSE}-free theory, the $\hat h$ terms in Eqs.~\eqref{eq:1eA_S} and~\eqref{eq:1eA_S2} (denoted as BSSE) as well as \eqref{eq:2eA_S} and \eqref{eq:2eA_S2} must not be taken into account as they are intramolecular, while similar terms \eqref{eq:2eAB_noS}--\eqref{eq:2eAB_S2} should stay as they are intermolecular.
For analysis purposes, we provide further discussion of  these \ac{BSSE} terms in the \supp; see Section S6.  

\subsection{Final expression}
\label{sec:final_expr}

Summarizing the results,  the final expression for the Hamiltonian matrix elements reads 
\begin{align}\label{eq:Ham_mat}
     H_{abcd} = &(E^A_a+E^B_b)\delta_{ac}\delta_{bd} \nonumber \\
     &+ \underbrace{V^{AB}_{ac}\delta_{bd} + V^{BA}_{bd}\delta_{ac} + J_{abcd}+V_\text{nn}\delta_{ac}\delta_{bd}}_\text{\elst}\nonumber\\
     &- \underbrace{K_{abcd} + X^{S}_{abcd}}_{\exch}  +N_{abcd} \, .
\end{align}
As before $E^M_i$ are monomeric state energies, Eq.~\eqref{eq:energy}, $V^{MN}$ are electron-nuclear coupling terms from Eq.~\eqref{eq:1eA_noS}, $J_{abcd}$ is the first term in Eq.~\eqref{eq:2eAB_noS} describing the interaction of electronic (transition) densities, and $V^{AB}_{ac}$ includes the interaction of the static electron density ($a=c$) or the transition density ($a\neq c$) of the monomer $A$ with the nuclei of $B$ and vice versa for $V^{BA}_{bd}$. The  regular exchange contribution $K$ is defined as 
\begin{align}\label{eq:K}
    K_{abcd}=-\sum_{\substack{pr\in A\\ qs \in B}}\gamma^{ac}_{pr}\gamma^{bd}_{qs}\braket{\orba{p}\orbb{q}|\orbb{s}\orba{r}} \nonumber \\
    -\sum_{\substack{kr\in A\\ ls \in B}}\gamma^{ac}_{kr}\gamma^{bd}_{ls} \braket{\orbb{l}\orba{k}|\orba{r}\orbb{s}}\, .
\end{align}
Finally, $X^{\text{S}}_{abcd}$ comprises all overlap-dependent exchange terms from Eq.~\eqref{eq:Vnn} and Eqs.~\eqref{eq:2eAB_S}--\eqref{eq:2eAB_S2}, non-\ac{BSSE} terms from Eqs.~\eqref{eq:1eA_S} and~\eqref{eq:1eA_S2},  as well as the corresponding parts from $\bra{\wf{A}{a}\wf{B}{b}}\P \hat H_{\text{inter}}\ket{\wf{A}{c}\wf{B}{d}}$. 
The lengthy expression is given in the \supp, Sec.~S1. 
Note that $J$, $K$, and \Sexch have diagonal and off-diagonal contributions to the Hamiltonian matrix. 
$N_{abcd}$ is the normalization correction, Eq.~\eqref{eq:normcorr}.

If we include exchange at the level of taking $\P$ into account, then besides the regular exchange contribution $K$, a number of additional terms appear that are comprised in \Sexch. 
Notice that $K$ is the  only two-electron exchange term that is dependent on one-electron (transition) densities. All  two-electron terms in \Sexch contain densities of higher particle order. Further, only $K$ does not depend on overlap integrals.  Methods such as the one by Pite\v{s}a et al.~\cite{Pitesa_JCTC_2024} or the approach based on the linear-response \ac{TDDFT} in Ref.~\citenum{Russo_JPCB_2007} involving the density functional to account for exchange take into account only $K$. In Section \ref{sec:comparison}  we will scrutinize the effect of \Sexch for a number of test systems.

Note that if the effect of exchange is neglected altogether, only \elst-coupling is considered.  This approximation should hold for moderate to large distances between monomers, i.e.,  cases where dipole-dipole (or transition density) interactions prevail.  We have already implemented it in \texttt{OpenMolcas} previously.~\cite{Kaiser_JCTC_2023}

Finally, we briefly comment on the calculation of eigenstates of the aggregate Hamiltonian, although we do not provide such calculations in this publication. 
The excitonic eigenstate vectors $\pmb{\Psi}$ and energies can be obtained as a result of solving a generalized eigenvalue problem. 
\begin{equation}\label{eq:gen_eigv}
    \vec H_\text{agg} \pmb{\Psi} = \vec E \vec\Sigma \pmb{\Psi} \, ,
\end{equation}
with the matrix elements of $ \vec H_\text{agg}$ given by Eq. \eqref{eq:Ham_mat} and $\vec \Sigma$ being the overlap matrix between basis excitonic states, Eq.~\eqref{eq:ovlp_neutral}.
However, a straightforward solution of the eigenvalue problem might mix different orders in the overlap and thus $\P$,~\cite{valeev06_9882} due to the presence of $\vec \Sigma$ matrix on the right-hand side of Eq.~\eqref{eq:gen_eigv}, with the consequences for the result being difficult to predict. 
To maintain a clear order in the $\P$ expansion, one has to perform orthogonalization, e.g., the symmetric one, 
\begin{equation}\label{eq:Loewdin_eigv}
    \vec \Sigma^{-1/2} \vec H_\text{agg} \vec \Sigma^{-1/2} \pmb{\Psi} = \vec E \pmb{\Psi} \, ,
\end{equation}
and sort out the respective terms; see \supp Sec.~S2. 
In the applications below, we will not discuss the eigenvalue problem further and focus on the discussion of the matrix elements of the aggregate Hamiltonian.

{In principle, one can also orthogonalize orbitals of the two monomers such that $\vec S^{AB}=\vec 0$.
It allows one to consider only the $K$ exchange term, as all the terms in $X^\text{S}$, $N$, and additional terms due to Eq.~\eqref{eq:Loewdin_eigv} (see Sec.~S2), explicitly depending on orbital overlaps, vanish.
It, however, requires the transformation of integrals and density matrices to a new orthogonal basis.}
Thus at the cost of performing such a transformation,
one effectively redistributes the respective exchange contributions to other terms, Sec.~S2.
However, we prefer to discuss overlap-dependent contributions explicitly, to emphasize the necessity of additional terms when working in a non-orthogonal basis.

\subsection{Relation to \ac{SAPT}}
\label{sec:comparison}
%
In what follows, we compare  the present approach to \ac{SAPT},~\cite{Jeziorski_CR_1994,Hapka_JCTC_2021,Szalewicz_WCMS_2012,Jansen_WCMS_2014,Patkowski_WCMS_2020,Patkowski_JMST_2001,Patkowski_JCP_2004} which  
is one of the most efficient approaches to describe intermolecular interactions in ground-state molecular dimers. In terms of excited states, the approach has been used, e.g., to study dispersion interactions in molecular dimers.\cite{hapka23_6895}
In \ac{SAPT}, one makes a double perturbation expansion with respect to intramonomer correlation and intermonomer interaction, usually with respect to a Hartree-Fock single-reference state.
Remarkably, the terms obtained in our approach that include exchange (\Sexch in Eq.~\eqref{eq:Ham_mat}) correspond to the first-order exchange correction in \ac{SAPT}, also called exchange repulsion.~\cite{Hapka_JCTC_2021} Note that applying the CHA approximation is vital for this correspondence as SAPT is BSSE free by construction. In the \supp\ {Sec.~S5}, we provide a numerical 
comparison with recent results obtained by 
Hapka et al.~\cite{Hapka_JCTC_2021} for the T-shaped  H$_{2}$-H$_{2}$ dimer and the C$_{2}$H$_{4}$--Ar excimer. 

However, it is important to realize that the present approach is different in principle.  
Specifically, we are using a generalization of the variational principle, Eq.~\eqref{eq:gen_eigv}, which is possible since the Hamiltonian is Hermitian, as we have deleted non-Hermitian \ac{BSSE} 
terms;~\cite{Mayer_IJQC_1998} it has a real spectrum, and the basis state overlap matrix $\vec \Sigma$, Eq.~\eqref{eq:ovlp_neutral}, is positive definite as long as the intermonomer distance is not too short.
The respective eigenfunctions can be expressed in a \ac{CI}-like manner via tensor products of monomeric wave functions $\Psi=\sum_{IJ} C_{IJ}\Psi^A_I\otimes\Psi^B_J$, which is also common for ``exact'' excitonic methods, e.g., active space decomposition.~\cite{Parker_JCP_2013}

The here used \ac{CI}-like treatment is beneficial compared to \ac{SAPT} as there are no problems with the definition of a Hermitian zero-order Hamiltonian for non-orthogonal excitonic states $\mathcal{A}\{\Psi_a\Psi_b\}$, which is a prerequisite for the construction of perturbation series.
The nonorthogonality naturally blends into our formalism, Eq.~\eqref{eq:gen_eigv}.
The same applies to the possible appearance of (quasi-)degenerate states, which require special attention when using perturbation theory.
Furthermore, our method is symmetric in contrast to the symmetry-adapted ones according to the classification of Stone.~\cite{Stone2013}
In other words, from the beginning, we work with the antisymmetrized wave functions, Eq.~\eqref{eq:antisym_WF}, instead of implying permutation spin symmetry only when computing energy corrections as in \ac{SAPT}.

One should note that, in contrast to the \ac{SAPT} literature, we will avoid classifying terms into induction and dispersion, thus retaining only the subdivision into electrostatics and exchange. 
The reason is that induction
\[
V^\text{ind}\sim\sum_c\bra{\wf{A}{a}\wf{B}{b}}\hat H\ket{\wf{A}{a}\wf{B}{c}}\bra{\wf{A}{a}\wf{B}{c}}\hat H\ket{\wf{A}{a}\wf{B}{b}}
\]
and dispersion
\[
V^\text{disp}\sim\sum_{cd}\bra{\wf{A}{a}\wf{B}{b}}\hat H\ket{\wf{A}{c}\wf{B}{d}}\bra{\wf{A}{c}\wf{B}{d}}\hat H\ket{\wf{A}{a}\wf{B}{b}}
\] 
contributions to the state energies, in our case,  are accounted for automatically in the course of the diagonalization of the aggregate Hamiltonian matrix.
As a side note, if both $a$ and $b$ are ground states of respective monomers, one might also wish to exclude their couplings to the excited ones to ensure size consistency, see the discussion in the Supporting Info of Ref.~\citenum{Li_JCTC_2017}.

Finally, it should be noted that both the  present method and \ac{SAPT} will be challenged once higher-order permutations need to be included, e.g., for tightly packed dimers. 
In fact, the truncation of the antisymmetrizer, Eq.~\eqref{eq:antisym_WF}, at the $\P$ term and neglect of charge-transfer effects are the most severe approximations employed in our formalism which both make it unreliable for close distances between monomers.
For instance,  the physical states complying with the Pauli principle are underlain by the continuum of non-physical ones where the fractional occupation number of an orbital can become more than two due to transfer from the other monomer. By including only the $\P$ term in the antisymmetrization, one incompletely projects out the nonphysical non-Pauli contributions which can lead to severe problems at short distances, especially for one-electron cross-monomer electron-nuclear attraction terms $V_\text{en}^{MN}$.  Such terms may require additional regularization.~\cite{Patkowski_JMST_2001,Patkowski_JCP_2004,Adams_TCA_2002}
%
\subsection{\texttt{OpenMolcas} implementation}
\label{sec:protocol}

The calculation of  the Coulomb couplings, \elst in Eq.~\eqref{eq:Ham_mat}, the direct exchange term, $K$ ($\propto S^0$), 
and further terms due to the $\P$ operator, \Sexch  (having $S^1$ and $S^2$ dependence), are implemented in the \texttt{RASSI} module of the \texttt{OpenMolcas} package.~\cite{LiManni_JCTC_2023}
These levels of approximation correspond to the keywords \texttt{EXCItonics} 
and \texttt{P1EXchange}.
The procedure and input philosophy are the same as detailed in our previous article.~\cite{Kaiser_JCTC_2023} 

In a nutshell, the calculation starts with the computation of the one- and two-electron integrals for a dimer. 
It is followed by two separate calculations for each monomer and two calls to the \texttt{RASSI} module.
The first call computes intermediates, which depend only on the orbital indices of the monomer $A$, and the second call complements these intermediates with the counterparts $B$ to obtain the respective matrix elements.
In other words, the sums in Eqs.~\eqref{eq:Vnn}-\eqref{eq:2eAB_S2} are decoupled into $A$- and $B$-dependent parts to increase efficiency.
The computational workflow then coincides with that depicted in Fig.~2 of Ref.~\citenum{Kaiser_JCTC_2023}.

To speed up the calculation of couplings, we make use
of the Cholesky decomposition as implemented in \texttt{OpenMolcas};~\cite{Aquilante_MP_2017,Aquilante_LTiCCaP_2011} see Ref.~\citenum{Kaiser_JCTC_2023}.
The computation of particular terms is done in the \ac{AO} basis for \elst and \ac{MO} basis for all other terms.
Further speed-up is achieved by using symmetry to obtain the wave functions of monomers.
However, note that the symmetry of the dimer can be much lower than that of isolated monomers.
To avoid complex and numerous symmetry correlation diagrams for each combination of monomers' and dimer's symmetry, the excitonic coupling is computed without symmetry considerations.
For all tested cases, the results with and without symmetry for monomers coincided within the numerical noise threshold.

In the current version, solvation effects due to solvent screening can be taken into account only at the monomer level using the polarized continuum model.
Proper dimer solvation is yet to be implemented.

\subsection{Computational Details}
\label{sec:comp_details}

\subsubsection{Monomer and dimer geometries}\label{sec:geoms}

For all test systems apart from butadiene, the monomer geometries were optimized at the B3LYP/6-31G(d) level with the \texttt{QChem 6.0} package.~\cite{QChem} 
The \ac{FOD} molecule was assumed to be planar for testing purposes. The $D_{2h}$ point symmetry group was used for polyacenes. 
The geometry of butadiene was taken from Ref. \citenum{Shu_JACS_2017} and is the equilibrium geometry from experiment.
For dimers of  \ac{FOD}, butadiene, and polyacenes, the coplanar orientation was used.
The molecules were placed in the $xy$ plane and the distance $\vec R_{AB}$ along the $z$ axis was varied from 2 to 4.5\,\AA\ {with a step of 0.1\,\AA}, Fig.~\ref{fig:systems} b).  Notice that these dimer geometries are chosen for simplicity and are not intended to present any real structural motif of, for instance, a molecular crystal.  Indeed, it is well-known that polyacenes such as tetracene crystallize in a herringbone structure. However, in the context of device performance, there has been substantial interest in substituted tetracenes, which show, e.g., (slightly) slip-stacked structures.~\cite{chi08_234} 

\subsubsection{Multi-reference calculations}

A multistate \ac{CASPT2} correction for dynamic correlation was applied on top of \ac{CASSCF} with the IPEA shift of 0.25 Hartree. 
A diffuse basis set was chosen to study the effects of intermonomer orbital overlap and exchange on excitonic couplings.
Diffuse functions were shown to be of importance for $\pi$-conjugated systems for predicting the right order of excited states~\cite{Dong_JCTC_2019} and excimeric properties of aggregates.~\cite{Krueger_IJQC_2019a}
Here, we systematically use a pragmatic Pople-type basis 6-311++G(d,p) unless otherwise stated. 
In addition, the basis set dependence of individual contributions to the total exchange coupling \exch was analyzed in Sec.~S7 of \supp. No significant changes in the individual coupling terms or in the total coupling were observed upon basis set increase.
All \ac{CASSCF}/\ac{CASPT2} calculations were performed with the developer version based on versions 23 and 24 of the \texttt{OpenMolcas} program package.~\cite{LiManni_JCTC_2023}
The contributions of different terms to total excitonic coupling were also compared to the results of the \ac{TDDFT} approach implemented in \texttt{Gaussian 16};~\cite{Russo_JPCB_2007} see Section~\ref{sec:tddft}.
The detailed analysis of the wave functions of the considered states is given in \supp\ Sec.~S3.

\paragraph*{FOD} 
In the case of \ac{FOD}, 14 electrons were distributed on 10 orbitals (CAS(14,10)) within a state-averaged procedure that included the 10 lowest electronic states without symmetry restrictions.
The active space comprised $\sigma$ and $\pi$ bonding and antibonding orbitals, as well as lone pairs of nitrogen and oxygen atoms; see Fig. S1 in \supp. 
Importantly, it also included a Rydberg orbital (LUMO+2 in Fig. S1), representing a combination of $3s$ and $3p$ \acp{AO} of nitrogen and oxygen,
which allowed us to obtain transitions comparable to the \ac{TDDFT} approach for the first three S$_1$--S$_3$ excited states. Note that  the S$_2$ and S$_3$ states involve excitation to the Rydberg orbital and therefore can be assigned to Rydberg transitions.
To study the effect of the completeness and diffuseness of the basis, additional calculations have been done for \ac{FOD}.
The corresponding results are shown in Sec.~S7.

\paragraph*{Butadiene}
For butadiene, 4 electrons were distributed on 4 $\pi$ and $\pi^\ast$ orbitals (CAS(4,4)) 
averaging over 3 states. 
The \ac{CASPT2} correction was computed with an imaginary shift of 0.2 Hartree. Although butadiene has the C$_{2h}$ symmetry, for the sake of simplicity, no symmetry was used for the calculation. However, the results do not differ from references that have used symmetry.~\cite{Shu_JACS_2017, Dong_JCTC_2019}

\paragraph*{Benzene and tetracene}
For benzene, all $\pi$ and $\pi^\ast$ orbitals were included in the active space, i.e., a CAS(6,6) active space was used.
For tetracene, the two active spaces used in the literature (CAS(8,8) \cite{Sanchez-Mansilla_PCCP_2022} and CAS(12,12) \cite{Plasser_JCTC_2017}) were compared, and it was decided in favor of CAS(8,8) space due to lower computational effort and still very similar results to the larger one. 
The $D_{2h}$ point symmetry was used for both molecules. 
In benzene, the ground S$_0$ state of $A_g$ symmetry and two excited states S$_1$ and S$_2$ of $B_{3u}$ and $B_{2u}$ symmetry were computed. 
In tetracene, the three states were computed in the $A_g$, $B_{3u}$, and $B_{2u}$ representations, thus, corresponding to state-averaged \ac{CASSCF} and multi-state \ac{CASPT2} calculations.

\begin{table*}[tb]
 \centering
	\begin{tabular}{cccccccccc}
        & \multicolumn{2}{c}{\ac{FOD}} & \multicolumn{1}{c}{Butadiene} & \multicolumn{2}{c}{Benzene} & \multicolumn{2}{c}{Tetracene} \\
         S$_0\rightarrow$  & \ac{CASPT2} &  \ac{TDDFT} & \ac{CASPT2}  & \ac{CASPT2} &  \ac{TDDFT} & \ac{CASPT2} &  \ac{TDDFT}  \\
        

        $\text{S}_1$ & 5.26 (5.9 $\cdot 10^{-3}$) & 5.17 (2.0 $\cdot 10^{-4}$)& 6.18 (1.1)& 4.84 (--)&  5.46 (--) & 3.0 (2.0 $\cdot 10^{-1}$)& 2.77 (7.7 $\cdot 10^{-2}$)  \\
        
        $\text{S}_2$ & 5.94 (2.6 $\cdot 10^{-3}$)&  5.90 (3.5 $\cdot 10^{-3}$) & 6.69 (--)&   6.48 (--)&  6.15 (--) & 3.4 (9.7 $\cdot 10^{-4}$)&  3.63 (1.8 $\cdot 10^{-3}$) \\

        $\text{S}_3$ & 6.45 (7.9 $\cdot 10^{-5}$)  &  6.23 (1.5 $\cdot 10^{-3}$) & --  & -- & -- & -- & -- \\
        
        $\text{S}_7$ & 11.65 (6.6 $\cdot 10^{-3}$)  &  --  & -- & -- & -- & -- & -- \\

    
        \end{tabular}
 \caption{
Comparison of the energies of the first electronic transitions (in eV) for all molecules calculated using CASPT2 and \ac{TDDFT} with the CAM-B3LYP functional. The basis set is in both cases 6-311++G(d,p).
In cases where symmetry was used, the symmetries of the respective states are also given for reference. 
}
\label{tab:transitions}
\end{table*}

\subsubsection{\ac{TDDFT} calculations}
\label{sec:tddft}

To verify consistency and test the applicability limits of the single-reference approaches, for \ac{FOD}, benzene, and tetracene, a comparison is done with the \ac{TDDFT} approach~\cite{Russo_JPCB_2007} as implemented in \texttt{Gaussian 16}~\cite{Russo_JPCB_2007}, employing the range-separated CAM-B3LYP functional~\cite{Yanai_CPL_2004} and the same 6-311++G(d,p) basis as for \ac{CASPT2} calculations. The approach of Ref.~\citenum{Russo_JPCB_2007} is a hybrid of the linear-response method~\cite{Hsu_JCP_2001, Iozzi_JCP_2004} and the perturbative treatment~\cite{Harcourt_JCP_1994} including contributions from  charge-transfer configurations.
Here, we are interested in comparing our results against Coulomb and exchange parts following from the \ac{TDDFT} method and avoid comparison of the contributions due to charge-transfer configurations as they are not entering our expressions.
In the \ac{TDDFT} approach, the coupling is computed from matrix elements of  
\begin{equation}
    V=\int \mathrm d \vec r \mathrm d \vec r' \gamma^{A\dagger}(\vec r)\left[\frac{1}{|\vec r - \vec r'|}+g_\text{XC}(\vec r, \vec r')\right]\gamma^B(\vec r') \, ,
\end{equation}
where $|\vec r- \vec r'|^{-1}$ is the direct Coulomb term and the \ac{DFT} response kernel $g_\text{XC}=g_\text{X}^\text{exact} + g_\text{C}^\text{DFT} + g_\text{X}^\text{DFT}$ consists of the exact non-local orbital-dependent exchange and the correlation and exchange terms dependent on local density.
Below, we will compare, besides the Coulomb term, only the contributions from the long-range exact exchange $g_\text{X}^\text{exact}$.
{This is justified by the fact that intermonomer interaction is long-range in nature as well. However, we assess the influence of the local \ac{DFT} exchange in \supp, Sec.~S4; in most cases, the contribution of local DFT exchange is minor.}
The monomer excitation energies of all systems, including CASPT2 and TDDFT results, are compiled in Tab. \ref{tab:transitions}.

\begin{figure*}[tb!]
\centering
  \includegraphics[width=0.9\textwidth]{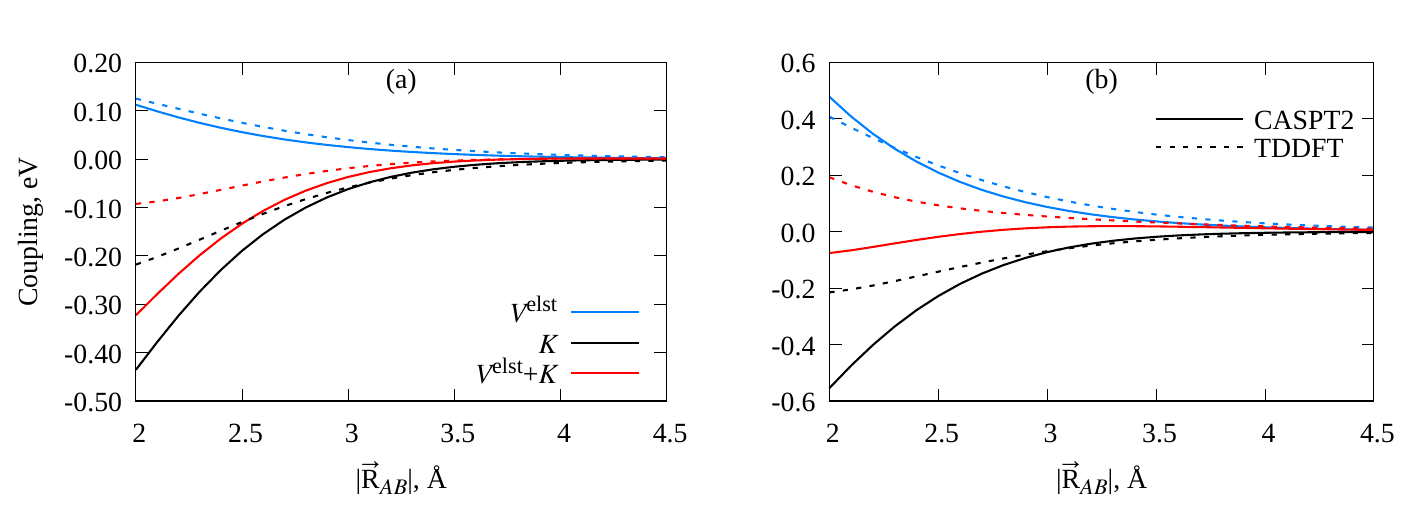}
  \caption{Benzene dimer: Distance dependence of the \elst and regular exchange $K$ (Eq.~\eqref{eq:Ham_mat}) couplings   using \ac{TDDFT} (dashed)  and \ac{CASPT2} (solid) methods. (a) S$_0\rightarrow \text{S}_1$ transition; (b) S$_0\rightarrow \text{S}_2$ transition. 
  \label{fgr:benz_jk}}
\end{figure*}

\begin{figure*}[tb!]
\centering
  \includegraphics[width=0.9\textwidth]{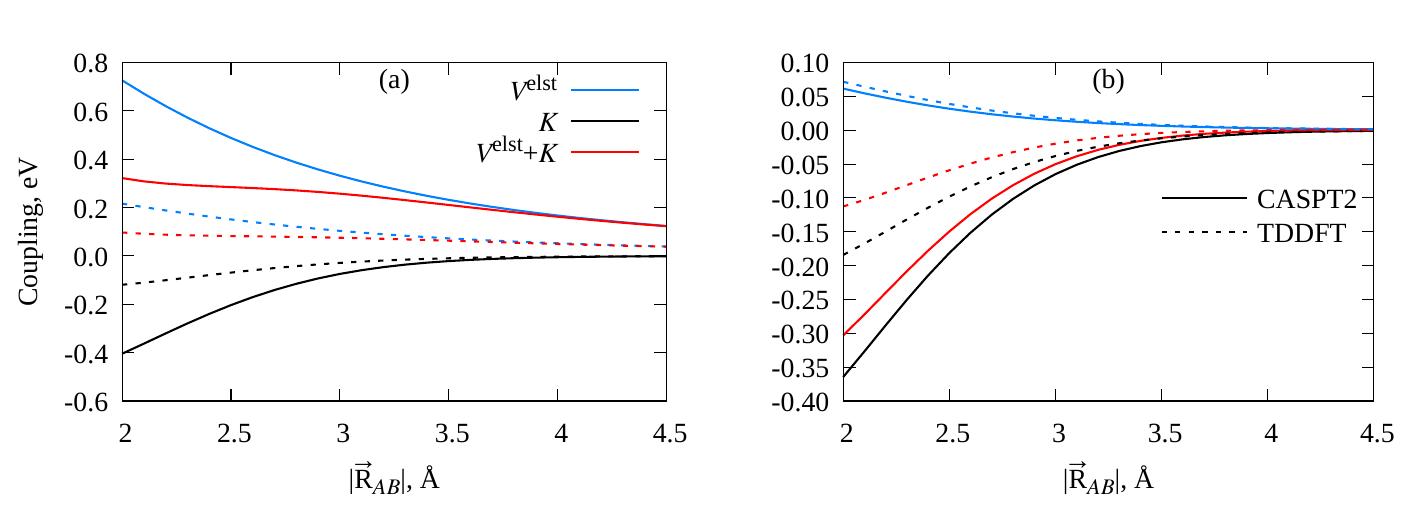}
  \caption{Tetracene dimer: Distance dependence of the \elst and $K$ couplings using \ac{TDDFT} and \ac{CASPT2} methods. (a)  S$_0\rightarrow \text{S}_1$ transition; (b)  S$_0\rightarrow \text{S}_2$ transition.}
  \label{fgr:tetr_jk}
\end{figure*}

\section{Results and Discussion}
\label{sec:results}

\begin{figure}[h]
\centering
  \includegraphics[width=0.415\textwidth]{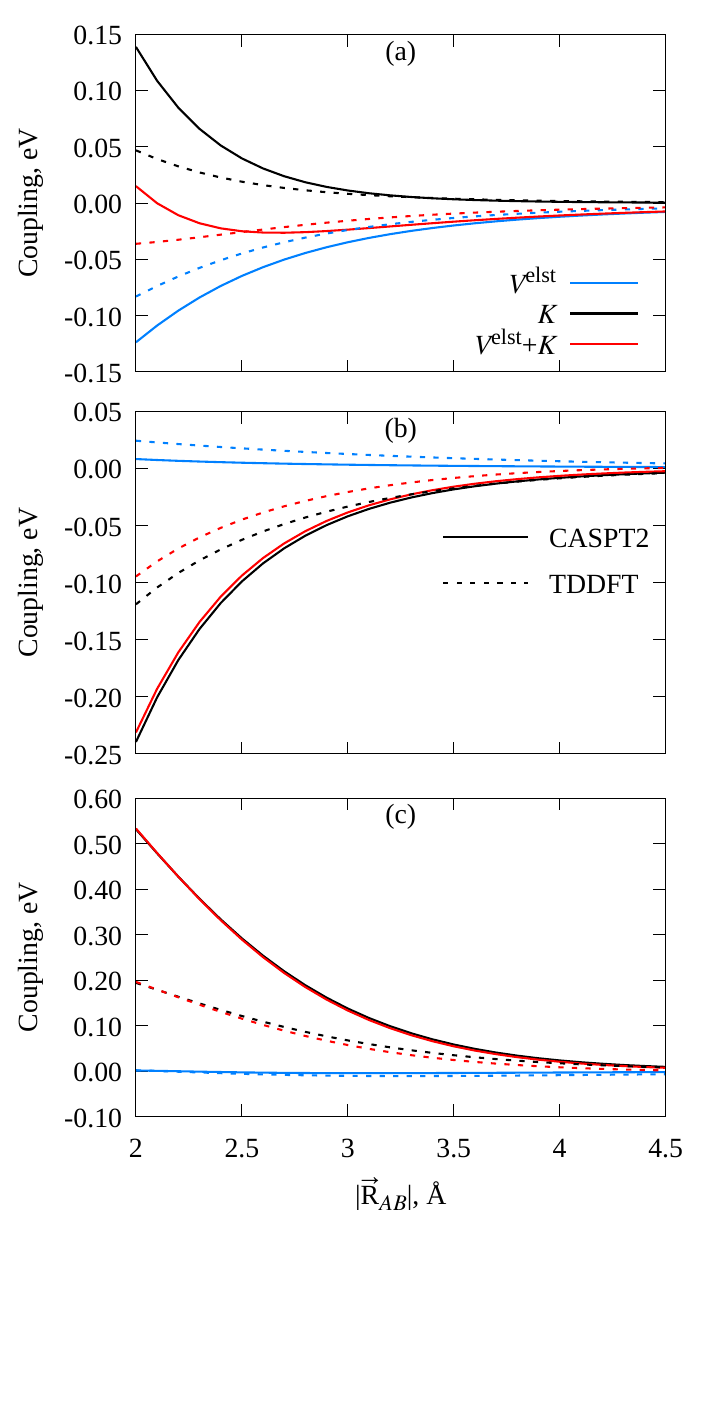}
  \caption{
  \ac{FOD} dimer: Distance dependence of the \elst and $K$ couplings using \ac{TDDFT} (dashed) and \ac{CASPT2} (solid) methods. 
    (a) S$_0\rightarrow \text{S}_1$,  (b) S$_0\rightarrow \text{S}_2$, (c) S$_0\rightarrow \text{S}_3$ transitions. }
  \label{fgr:fod_jk}
\end{figure}

\begin{figure}[tb!]
\centering
  \includegraphics[width=0.415\textwidth]{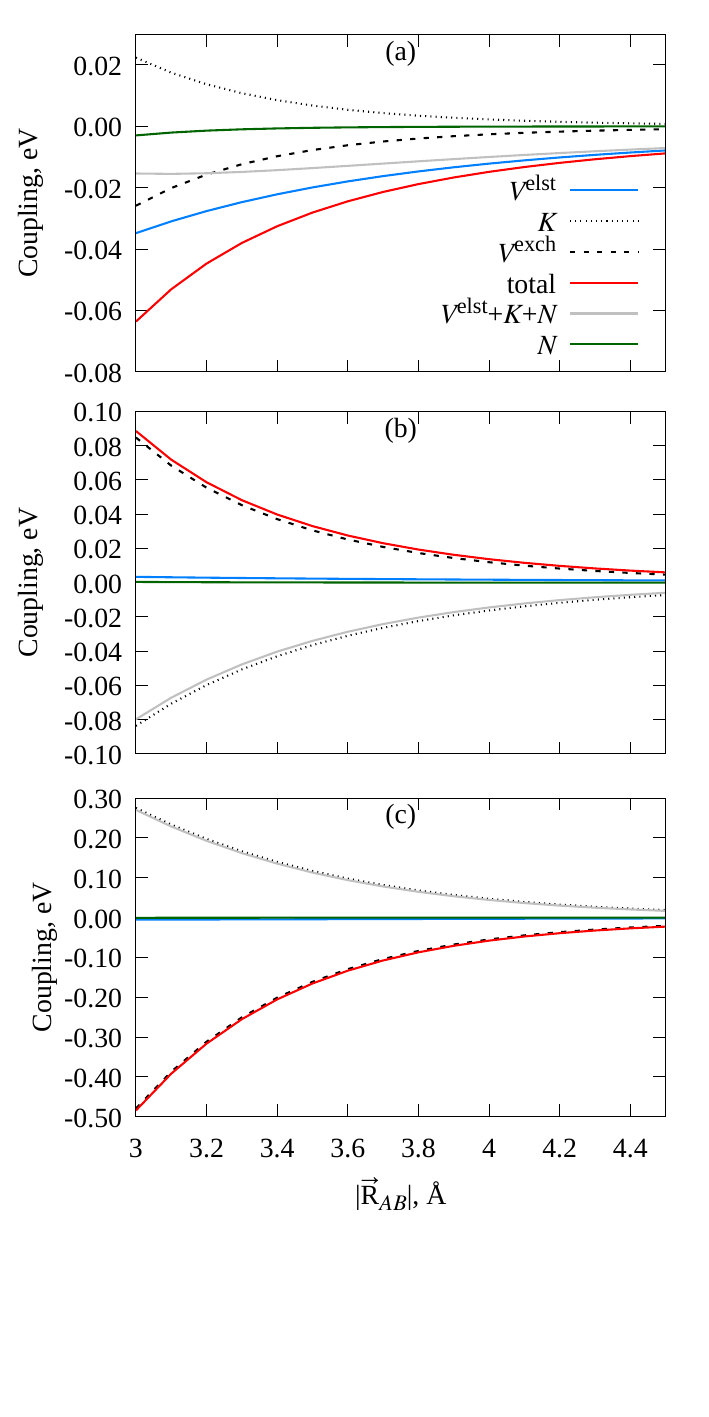}
  \caption{
  \ac{FOD} dimer: Coulomb (\elst), regular exchange ($K$), and full exchange (\exch) couplings for three lowest transitions.   (a) S$_0\rightarrow \text{S}_1$, (b) S$_0\rightarrow \text{S}_2$, (c) S$_0\rightarrow \text{S}_3$}
   \label{fgr:fod_sapt_k}
\end{figure}

\subsection{Regular exchange ($K$) for single- and multi-configurational cases}
\label{sec:regK}

Before focusing on the importance of the newly derived exchange terms proportional to overlap, we discuss the widely used regular exchange coupling $K$, the first term in Eq.~\eqref{eq:K}, which is not directly proportional to overlap, in conjunction with  \ac{TDDFT} in particular.~\cite{QChem,Russo_JPCB_2007}
We have chosen three systems, each illustrating different aspects that might be critical when the dimer states are computed on the basis of  $K$ only.

\paragraph*{Benzene} 
%
We compare the results obtained with \ac{CASPT2} and \ac{TDDFT} methods for the first two electronic transitions (see Table~\ref{tab:transitions}); these results agree reasonably well with the literature~\cite{C.A.Valente_JCP_2021a,AguilarSuarez_MP_2020,Hashimoto_JCP_1996} when the corresponding method is applied.  
The nature of the states is analyzed in the \supp\ Sec.~S3, confirming that these are the same states obtained with different methods.
The oscillator strengths for both transitions are zero as they are forbidden by symmetry. 
The \ac{CASPT2} and \ac{TDDFT} transition energies notably deviate from each other, with the \ac{CASPT2} results being in good agreement with other calculations and experiments.~\cite{Christiansen_JCP_1996,Roos_CPL_1992}
In Fig.~\ref{fgr:benz_jk}, we show the electrostatic, \elst, and $K$ coupling for both methods. 
Despite the vanishing dipole-dipole interaction, the \elst coupling is nonzero because of the interaction of extended transition densities. 
Although benzene has a notably multi-configurational ground state (see \supp), the couplings for the single-reference \ac{TDDFT} and multi-reference \ac{CASPT2} agree very well for distances greater than 3\,\AA. 
Therefore, the difference in wave functions, which is reflected in the notable differences in diagonal state energies, does not necessarily lead to significant differences in couplings.

\begin{figure*}[tb!]
\centering
  \includegraphics[width=0.9\textwidth]{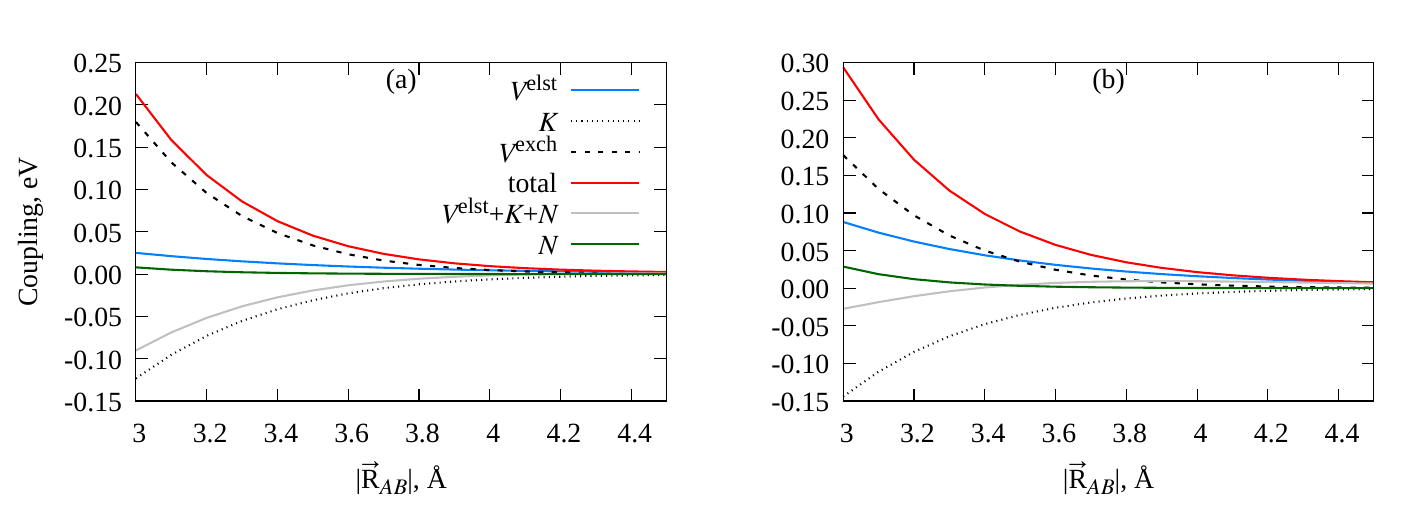}
  \caption{Benzene dimer: Distance dependence of the \elst, $K$, and \exch. (a)  S$_0\rightarrow \text{S}_1$ transition; (b)  S$_0\rightarrow \text{S}_2$ transition. 
  }
  \label{fgr:benz_k_sapt}
\end{figure*}

\begin{figure*}[tb!]
\centering
  \includegraphics[width=0.9\textwidth]{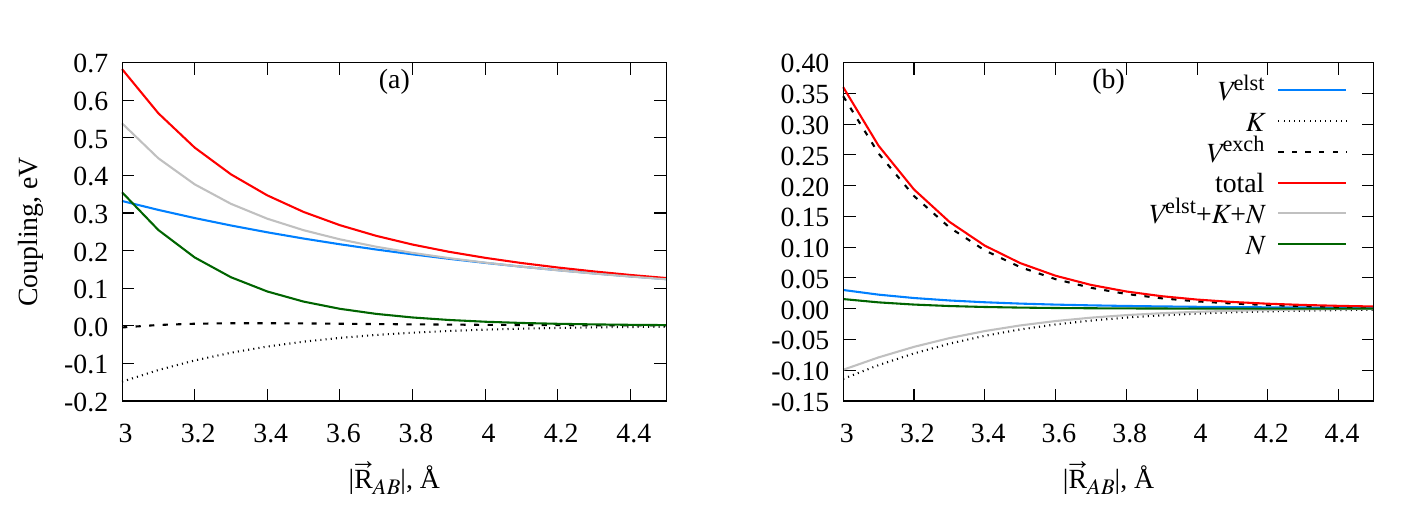}
  \caption{Tetracene dimer: Distance dependence of the \elst, $K$, and \exch. (a) S$_0\rightarrow \text{S}_1$ and (b)  S$_0\rightarrow \text{S}_2$ transition. 
  }
  \label{fgr:tetr_jk_exch}
\end{figure*}

\paragraph*{Tetracene}
Unlike benzene, tetracene displays the lowest electronic transition with notable oscillator strength.\cite{Wang_JMCA_2023} Excitation energies (Tab.~\ref{tab:transitions}) and oscillator strengths (Sec.~S3) computed with \ac{TDDFT} and \ac{CASPT2} agree well with each other for both S$_0\rightarrow $S$_1$ and S$_0\rightarrow $S$_2$ transitions.
Despite this agreement, the coupling strengths strongly deviate.
Fig.~\ref{fgr:tetr_jk} indicates a substantial difference of \ac{TDDFT} and \ac{CASPT2} \elst term for the first transition, while the second transition remains in accordance, as shown in panels (a) and (b), respectively. 
This is noteworthy since such a discrepancy for one transition and agreement for the other cannot be explained by the strength of the dipole-dipole interaction, as it should give notable differences in both cases (\supp). 
Such discrepancies might be explained by the differences in the outer parts of the transition densities for the two methods.
Although the general shapes of the orbitals and main electronic configurations can be unambiguously identified to directly compare the results of the methods if $D_{2h}$ symmetry is imposed, the notable multi-configurational nature of the tetracene wave functions leads to notable differences when \ac{CASPT2} results are compared to the single-reference \ac{TDDFT}.

\paragraph*{FOD}
The first three transitions show good agreement in terms of energies (Table~\ref{tab:transitions}) and state character (see \supp) between the \ac{CASPT2} and \ac{TDDFT} methods. 
The second and third excited states have Rydberg character, with excitation occurring to a mixture of $3s$ and $3p$ orbitals of N and O atoms. 
This is a critical characteristic as \ac{TDDFT} is known to poorly predict the properties of Rydberg states, with range-separated functionals partially mitigating this drawback.~\cite{Hirata1999,Hirata_CPL_1999b,Tozer_PCCP_2000,Casida_JCP_1998,Eriksen_MP_2013,Fuks_PRA_2014}
Whereas for the first transition one can see the correspondence between the \elst and $K$ coupling terms between the methods, for the other two the electrostatic contribution to the coupling in \ac{TDDFT} deviates significantly from the \ac{CASPT2} result.
Starting from distances >3\AA, the agreement becomes almost quantitative for both, the \elst and the $K$ couplings, except for the $K$ coupling of the S$_0\rightarrow \text{S}_3$ transition (see panel (c)), for both methods. The poor agreement of the $K$ coupling for the third S$_0\rightarrow \text{S}_3$ transition, shown in panel (c) of Fig.~\ref{fgr:fod_jk}, may be attributed to the Rydberg character of this transition and its very small oscillator strength.
Here, TDDFT fails to describe the Rydberg transition and strongly underestimates the coupling between these transitions despite the usage of the range-separated CAM-B3LYP functional.\\

In summary, we have identified situations where only the energies of the monomers are influenced by multi-reference effects, whereas the coupling stays similar in both single-reference, e.g., \ac{TDDFT}, and multi-reference methods. 
An example of such a system is the benzene dimer.
However, the opposite can also occur when the coupling is more sensitive to the quality of the wave function than the energy, as exemplified by tetracene and \ac{FOD}.
Remarkably, the $K$ term can be particularly strongly influenced by the Rydberg character of excitation.

\begin{figure*}[tb!]
\centering
  \includegraphics[width=0.9\textwidth]{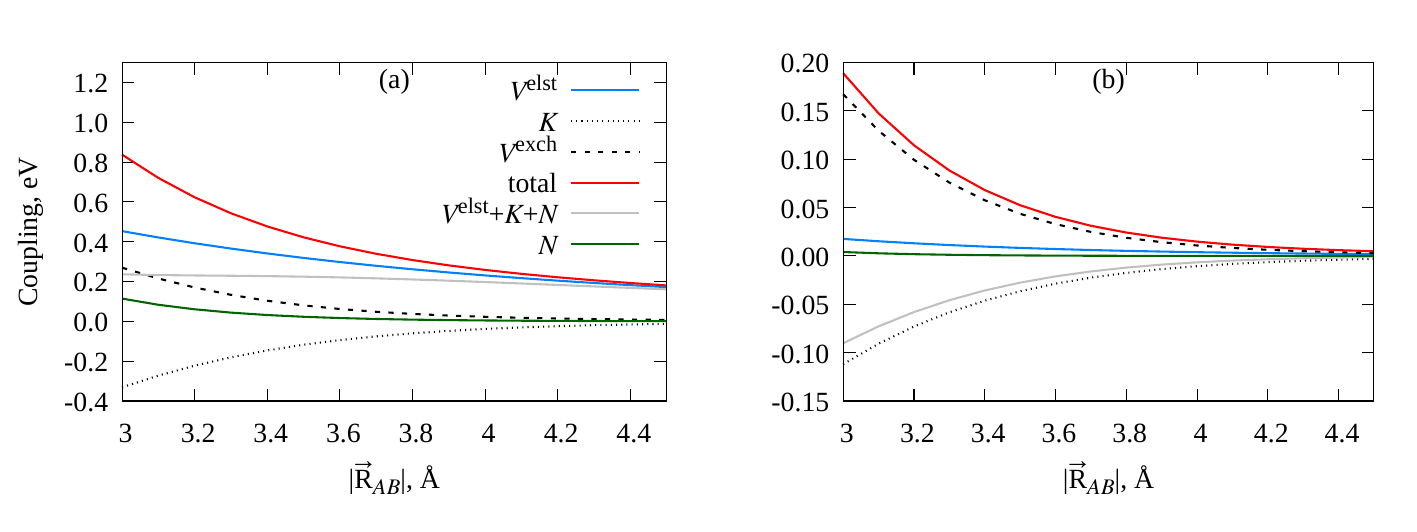}
  \caption{Butadiene dimer: Distance dependence of the \elst, $K$, and \exch couplings. (a) S$_0\rightarrow \text{S}_1$ and (b)  S$_0\rightarrow \text{S}_2$ transitions 
  }
  \label{fgr:buta_tdm}
\end{figure*}
\subsection{Role of overlap-dependent  exchange} 
\label{sec:exch_vs_k}
According to the derivation in Section~\ref{sec:theory}, the regular exchange  coupling term, $K$, is not the only exchange contribution: There are additional terms dependent on overlap and collected in the $X^\text{S}$ in Eq.~\eqref{eq:Ham_mat} that are equivalent to the first-order exchange correction to the interaction if viewed from the standpoint of \ac{SAPT}. 
Note that the individual contributions to $\exch$ are large and are subject to substantial mutual compensation, while \ac{BSSE} terms destroy such a balance; see \supp\ Sec.~S6.

In this section, we compare $K$, which is exclusively used in excitonic calculations instead of the full exchange (if any exchange effect is included at all),~\cite{QChem,Russo_JPCB_2007} with the total \exch$=K+X^\text{S}$.
In this context, we note that the shape of the dimer spectrum and the character of changes in aggregates with respect to non-interacting monomers depend on the sign of the coupling. 
One distinguishes between the formation of so-called J- (negative coupling, red shift) and H-aggregates (positive coupling, blue shift).\cite{may23}
In the following, we will focus on examples in which the type of aggregate changes depending on the level of inclusion of exchange effects.

\paragraph*{FOD} 
Figure \ref{fgr:fod_sapt_k} shows the distance dependence of the couplings for the three lowest electronic transitions of \ac{FOD}. 
In the case of the S$_0\rightarrow\text{S}_1$ transition, 
the Coulomb coupling dominates the exchange couplings; thus, the total coupling in both cases leads to a negative value.
The magnitude of \exch and $K$ is comparable, but the signs of \exch and $K$ are opposite. 
For the S$_0\rightarrow \text{S}_2$ transition, we observe different signs of $K$ and \exch, which are both notably larger than \elst. 
Taking into account only \elst or \elst+\exch gives a positive coupling leading to an H-aggregate.
However, if one takes into account only the $K$ term instead of the full \exch, one obtains a negative sign of the total coupling, implying the formation of a J-aggregate.
In the case of the S$_0\rightarrow\text{S}_3$ transition, the exchange terms also dominate electrostatic coupling for the considered distances, but the signs of $K$ and \exch coincide.
For all three cases, the norm correction ($N$) is negligible for the whole distance range investigated.

\paragraph*{Benzene} 

As expected, the electrostatic coupling is very small compared to \exch and $K$, due to the dark nature of the considered transitions; see Fig.~\ref{fgr:benz_k_sapt}. For both transitions, the signs of $K$ and \exch are opposite, while dominating \elst. Thus, again the choice of the exchange is of importance for the kind of aggregate formed. For the S$_0\rightarrow \text{S}_2$ transition, the opposite sign leads to a cancellation, i.e., the total coupling almost vanishes. Also here, the normalization factor is negligibly small.

\paragraph*{Tetracene} 
%

In panel (a) of Fig. \ref{fgr:tetr_jk_exch} (S$_0\rightarrow \text{S}_1$ ), one sees a negative sign of the $K$ coupling  and a dominating \elst coupling over the whole distance, while $\exch$ stays nearly zero. Noteworthy is the relatively large normalization constant, which did not play a role in the examples discussed so far. For the other transition (S$_0\rightarrow \text{S}_2$ ), panel (b), one again observes opposite contributions of $K$ and overlap-dependent terms, hinting that selecting $K$ as the sole exchange term can lead to qualitatively wrong results.\\

In summary, the above cases demonstrate the importance of including the full exchange term \exch in relation to total coupling, which can change the sign of the matrix elements. Of course, this conclusion hinges on the validity of the $P^{(1)}$ approximation, which will be further scrutinized in Sec. \ref{sec:validity_P1}.
However, note that the generalized eigenvalue problem Eq.~\eqref{eq:gen_eigv} introduces some complexity due to the non-orthogonality of the basis states. Nevertheless, in test cases for intermonomer distances above 3\AA, we have seen that the nature of the optical response of the aggregate (J- or H-type) has not changed upon solution of Eq.~\eqref{eq:gen_eigv}. 

\subsection{Doubly excited states} 
\label{sec:double_rydberg}

In this section, we will investigate the behavior of the different kinds of intermonomer couplings in cases where single-reference methods fail to describe the systems properly. Those cases are dark states that arise from double excitations and transition to Rydberg states.
For instance, linear-response \ac{TDDFT} or wave-function single-reference methods with insufficient level of excitation, e.g., equation-of-motion or linear-response methods including only up to double excitations, could be examples of techniques that may experience problems with such transitions, whereas multi-reference methods suit their description very well.

\paragraph*{Butadiene} 

The second excited state of butadiene is dark and has a significant double excitation character~\cite{Shu_JACS_2017}, while the first transition is bright; see \supp.
This is reflected in the strength of electrostatic coupling, which is an order of magnitude larger for the bright state, 
Fig.~\ref{fgr:buta_tdm}. As expected, the \elst coupling is negligibly small to the exchange couplings \exch  and $K$. Figure~\ref{fgr:buta_tdm} (b) also represents another example where the terms $K$ and \exch have opposite signs. Thus, whether the total coupling is positive or negative strongly depends on the chosen variant of exchange coupling. 
 
\begin{figure}[t]
\centering
  \includegraphics[width=0.45\textwidth]{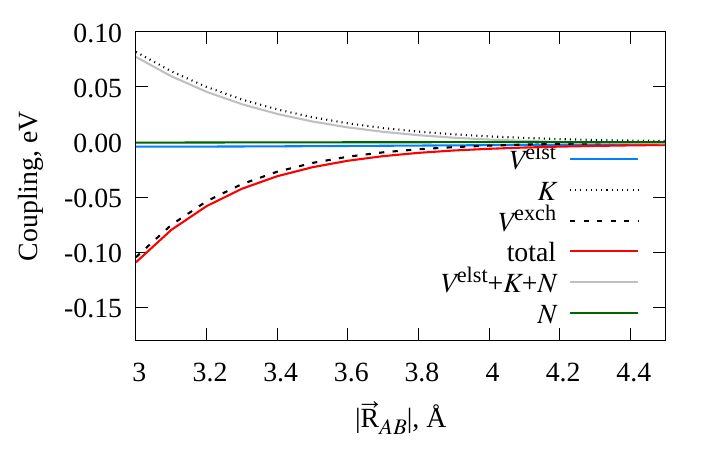}
  \caption{ Distance dependence of the \elst, $K$, and \exch couplings of the \ac{FOD} dimer for the S$_0\rightarrow \text{S}_7$ transitions.}
  \label{fgr:fod_1-7}
\end{figure}

\paragraph*{FOD} 
The S$_0\rightarrow\text{S}_7$ transition of \ac{FOD} has both a double-excitation and Rydberg character simultaneously; see \supp.   
The oscillator strength for this transition is relatively small ($\simeq$10$^{-3}$), and the electrostatic coupling is almost vanishing, Fig.~\ref{fgr:fod_1-7}. 
As for other transitions of \ac{FOD}, the exchange coupling is notable and even dominating for Rydberg states, where the orbital overlap is large.
Fig.~\ref{fgr:fod_1-7} also highlights the importance of selecting the appropriate exchange coupling since $K$ and \exch have opposite signs. 
 
\subsection{{Validity of the $\P$ approximation}}
\label{sec:validity_P1}

The overlap is the most basic measure of the significance of the exchange effect, as all terms rely on it, either directly (Eqs.~\eqref{eq:2eBA_S}, \eqref{eq:2eAB_S}, and~\eqref{eq:2eAB_S2}) or indirectly, such as $K$ in Eq.~\eqref{eq:2eAB_noS}. 
To scrutinize the range of validity of the  $P^{(1)}$ approximation, in what follows,  we will compare it to $P^{(2)}$ for the particular case of FOD. 

\begin{figure*}[tb!]
    \centering
    \includegraphics[width=0.9\linewidth]{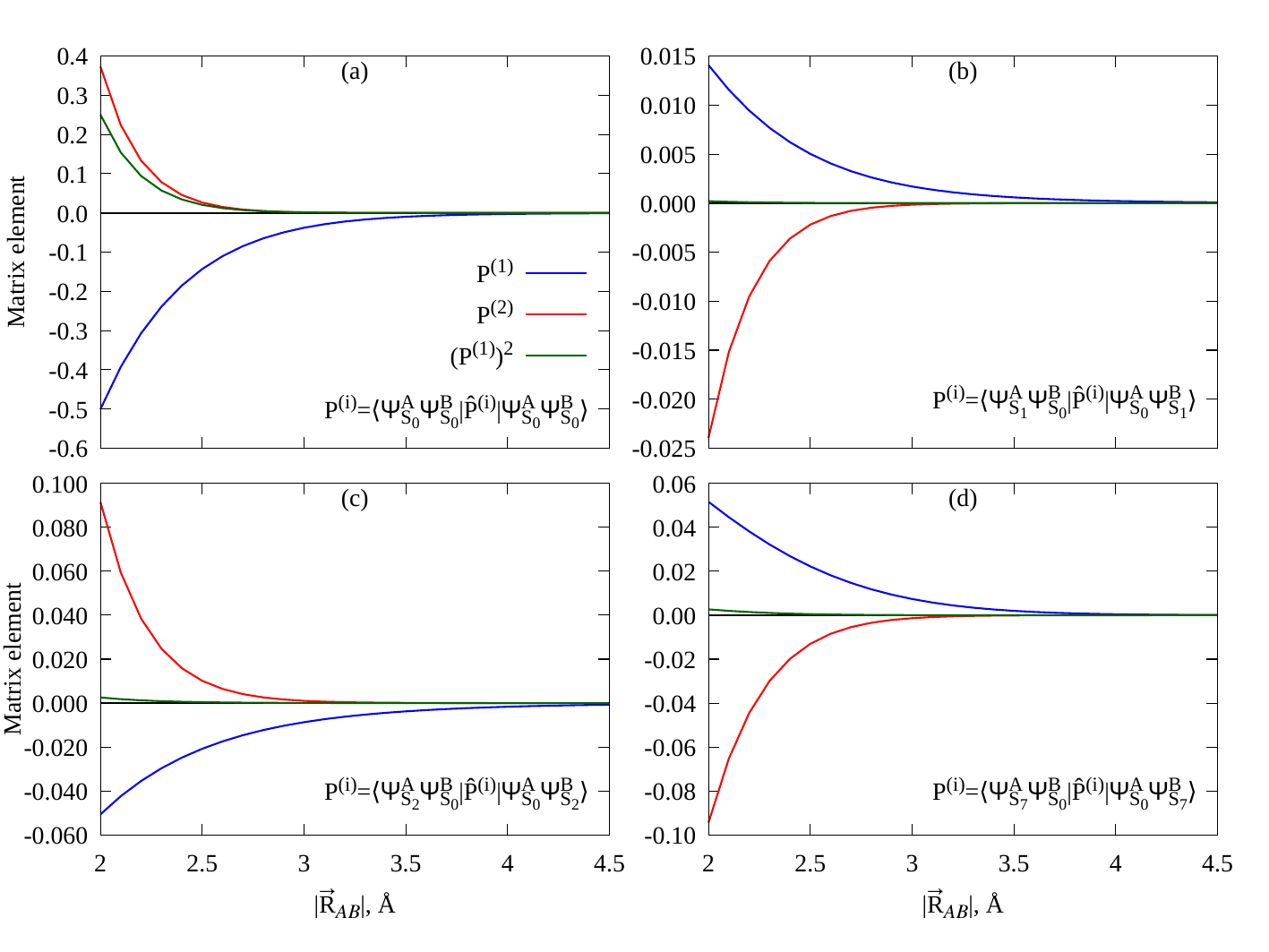}
    \caption{FOD dimer: Matrix elements $\braket{\wf{A}{S_0}\wf{B}{S_0}|\hat P^{(i)}|\wf{A}{S_0}\wf{B}{S_0}}$, $\braket{\wf{A}{S_1}\wf{B}{S_0}|\hat P^{(i)}|\wf{A}{S_0}\wf{B}{S_1}}$, $\braket{\wf{A}{S_2}\wf{B}{S_0}|\hat P^{(i)}|\wf{A}{S_0}\wf{B}{S_2}}$ and $\braket{\wf{A}{S_7}\wf{B}{S_0}|\hat P^{(i)}|\wf{A}{S_0}\wf{B}{S_7}}$, for $i=1,2$, in panels (a), (b), (c) and (d), respectively. Also, the square of the matrix elements of the $\P$ operator is shown for illustrative purposes. 
    }
    \label{fig:fod_ovlp_z}
\end{figure*}

The matrix elements $P^{(1)}$ of the operator Eq. \eqref{eq:P1} in the Frenkel basis read
\begin{align}\label{eq:P1_mtx}
     \braket{\wf{A}{a}\wf{B}{b}|\hat P^{(1)}|\wf{A}{c}\wf{B}{d}}=-\sum_{\substack{kn\in A\\lm\in B}}\sab{km}\sba{ln}\gamma^{ac}_{kn}\gamma^{bd}_{lm} \,.
\end{align}
The formal expression for the two-particle exchange operator is given by 
\begin{align}
    \hat P^{(2)} = \sum_{\substack{iktu\in A\\jlvx\in B}}&\sba{jt}\sba{lu}\sab{iv}\sab{kx}
\crea{t}\crea{u}\anna{i}\anna{k}\otimes\creb{v}\creb{x}\annb{j}\annb{l}\,.
\end{align}
Taking the respective matrix elements we obtain
\begin{align}\label{eq:P2}
     P^{(2)} &=\braket{\wf{A}{a}\wf{B}{b}|\hat P^{(2)}|\wf{A}{c}\wf{B}{d}} \nonumber \\
     &= \sum_{\substack{iktu\in A\\jlvx\in B}}\sba{jt}\sba{lu}\sab{iv}\sab{kx}\gamma_{tuik}^{ac}\gamma_{vxjl}^{bd}\,.
\end{align}
Figure \ref{fig:fod_ovlp_z} shows $P^{(1)}$, $P^{(2)}$, and $(P^{(1)})^2$ for FOD ground and transition matrix elements.
All curves show the expected exponential scaling with respect to distance with $P^{(1)} \propto S^2$ decaying slower than $P^{(2)} \propto S^4$. In addition the ground state matrix elements, $\braket{\wf{A}{S_0}\wf{B}{S_0}|\hat P^{(i)}|\wf{A}{S_0}\wf{B}{S_0}}$,  decay faster than those involving a local excitation.

The effective distance beyond which higher-order terms can be neglected is approximately 3\AA. Assuming that the FOD results are transferable to other systems, such a distance is below typical intermonomer distances in molecular crystals such as those formed by substituted tetracenes. \cite{chi08_234}
{It approximately corresponds to distances where charge-transfer effects also start to play a role.}

\section{Conclusion and Outlook}
\label{sec:conclusions}

In this study, we have derived and implemented approximations to the overlap-dependent exchange interaction between the monomeric transitions in molecular aggregates.
The derivation followed the interacting-monomers strategy of the Frenkel exciton Hamiltonian and employed the single electron pair exchange approximation.\cite{may23}
The primary objective of this work was to integrate this approach with multi-reference electronic structure methods, such as \ac{CASPT2}, and to evaluate the significance of exchange effects for molecules exhibiting notable multi-configurational wave functions in the ground and/or lowest excited electronic states.
To this end the approach was implemented in the \texttt{RASSI} module of the \texttt{OpenMolcas} software.\cite{LiManni_JCTC_2023}

To arrive at a physically meaningful coupling, the \acl{CHA}~\cite{Mayer_IJQC_1998} has been applied, thus excluding the \ac{BSSE} terms.
The treatment of exchange in our approach extends beyond the regular exchange term $K$, which is commonly included in theories accounting for exchange effects, and involves additional terms that depend on the linear or quadratic overlap of the monomer wave functions.
The resulting coupling terms coincide in their form with the first-order exchange repulsion of \ac{SAPT}. However, our approach is non-perturbative and is based on a variational principle.

We have provided illustrative examples where the couplings obtained by a single-reference method (\ac{TDDFT}) and those obtained with the present method align with each other, and where discrepancies arise, primarily due to the occurrence of multi-reference effects, double excitations, and the Rydberg character of transitions. 
Besides the accuracy of the electronic structure prediction, two prominent physical effects have been observed.

For Rydberg transitions, we found that the coupling is determined almost solely by the exchange effect. This is due to the combination of the smallness of the transition dipoles and pronounced overlap between diffuse Rydberg states. The dominance of exchange coupling is commonly expected for Dexter-type transfer between triplet states, but is observed here for singlet excitations.

Second, the inclusion of overlap-dependent exchange terms ($X^{\rm S}$) in some cases did give a qualitative difference as compared with the $ K$-only theory. Specifically, it was observed that the character of the dimer changes between J- and H-type, with all its profound consequences for the photophysics.

Finally, two potential limitations of the present approach need to be discussed. First, the single-electron exchange approximation. Here we observed for the specific case considered that due to the expansion in terms of powers of the wavefunction overlap higher order contributions become sizable for distances below about 3~\AA. This puts the present approach on the safe side for many molecular crystals and photosynthetic systems, such as the reaction centers' special chlorophyll pairs.
Second, at  distances between 3 and 4 \AA{}, charge transfer transitions often mix with local excitations. 
The present  theoretical framework is well-suited to incorporate coupling to charge transfer transitions. 

\section*{Conflicts of interest}
There are no conflicts to declare.


\section*{Acknowledgements}

The financial support from Deutsche Forschungsgemeinschaft Grants No. Ku952/10-2 (A.K. and O.K.) and 521855798 (S.I.B.) is gratefully acknowledged.

\bibliography{Excitons_exchange} 
\bibliographystyle{unsrtnat} 

\end{document}